\documentclass[fleqn,usenatbib,useAMS]{mnras}
\usepackage{graphicx}	
\usepackage{amsmath,mathtools}	
\usepackage{amssymb}	
\usepackage{multicol} 

\usepackage{bm}		
\usepackage{pdflscape}	

\usepackage{verbatim,xspace}
\usepackage{import}

\usepackage[T1]{fontenc}
\usepackage{ae,aecompl}
\usepackage{newtxtext,newtxmath}

\newcommand{\ro}[1]{\ensuremath{\textrm{#1}}}
\newcommand{\ten}[1]{\ensuremath{\times 10^{#1}}}
\newcommand{\sm}{\ensuremath{{\sim}}}

\newcommand{\kmsec}{\ensuremath{~\ro{km}~ \ro{s}^{-1}}\xspace}

\newcommand{\ergcm}{\ensuremath{~\ro{erg cm}^{-2}}}

\newcommand{\Msol}{\ensuremath{M_{\odot}}\xspace}
\newcommand{\Zsol}{\ensuremath{Z_{\odot}}\xspace}

\newcommand{\df}{\ensuremath{~ \ro{d} }}
\newcommand{\dd}{\ensuremath{\ro{d} }}

\newcommand{\eagle}{\textsc{eagle}\xspace}

\newcommand{\pcc}{\ensuremath{P_\ro{CC}}\xspace}
\newcommand{\pia}{\ensuremath{P_\ro{Ia}}\xspace}

\title[Life in the Cosmic Neighbourhood]{Life in the Cosmic Neighbourhood: Galactic Habitable Zones in the EAGLE Simulations} 

\author[L. Barnes et al.]{%
Luke A. Barnes,$^{1}$\thanks{E-mail: L.Barnes@westernsydney.edu.au} 
Geraint F. Lewis$^2$, Jaime Salcido$^{3,4}$, Pascal J. Elahi$^5$, \newauthor
Miroslav D. Filipovic$^1$, and Richard G. Bower.$^6$
\\ \\
$^1$ Western Sydney University, Faculty of Engineering, Computing and Science, Penrith, NSW, 2751, Australia \\
$^2$ Sydney Institute for Astronomy, School of Physics, A28, The University of Sydney, NSW 2006, Australia \\
$^3$ Visa Cash App Racing Bulls Formula One Team, Milton Keynes, England, UK\\
$^4$ Astrophysics Research Institute, Liverpool John Moores University, 146 Brownlow Hill, Liverpool L3 5RF, UK\\
$^5$ Pawsey Supercomputing Research Centre, 1 Bryce Avenue, Kensington WA, 6151, Australia\\
$^6$ Institute for Computational Cosmology, Department of Physics, University of Durham, South Road, Durham, DH1 3LE, UK
}

\date{Last updated Today; in original form Yesterday}

\pubyear{2026}

\hypersetup{draft}

\begin{document}
\label{firstpage}
\pagerange{\pageref{firstpage}--\pageref{lastpage}}
\maketitle

\begin{abstract}
When and where is life most likely to form in the universe? We use the \eagle cosmological hydrodynamical simulations to investigate the prevalence and properties of life-friendly cosmic environments on galactic scales. Our model for the occurrence of observers depends on main-sequence stars, the production and distribution of metals, and the formation of habitable planets. We also take into account five potentially life-extinguishing processes: core-collapse supernovae, Type Ia supernovae, star-star interactions, gamma ray bursts and quasars. We characterise the amount of time that the universe creates on habitable planets per unit mass of baryons and stars in the universe, as a function of galaxy mass and galactocentric distance. As expected, early universe, newly formed, high star formation rate galaxies are the most dangerous. The present-day Milky Way is typical of habitable galaxies: efficient creation of habitable time, and low probability of cosmic catastrophes.
\end{abstract}



\section{Introduction} 
Astrobiology seeks to characterise the suitability of various cosmic environments for life. As reviewed in \citet{2005OLEB...35..555G} and \citet{2012AREPS..40..597L}, there has been much discussion of the circumstellar habitable zone (CHZ), which characterises the likely place of life in planetary systems. Liquid water is very important, which implies that the planet must not be too close or too far from its host star. Further work has sought to model the effect of atmospheric loss, volcanic outgassing, biomass feedback, stellar variability, sources of internal heat, cloud formation, planetary migration and more.

Life also depends on its larger cosmic environment, which has lead to the study of the \emph{galactic habitable zone} (GHZ). Again, a variety of important factors are at play. The formation of planets depends on the abundance of metals in the stellar nurseries. Once a planet has formed, its habitability will be affected by energetic nearby events such as supernovae (SN), gamma ray bursts (GRBs) and quasars, particularly if the star is born in a cluster or in the dense core of a galaxy. The study of energetic phenomena on cosmic habitability is termed ``high-energy astrobiology'' in the edited volume of Filipovic (forthcoming).

Study of the GHZ has overlapped with the study of the effect of our cosmic environment on the actual history of Earth. For example, can any of Earth's mass extinctions be explained by the effect of a nearby SN or GRB? How close does such an explosion need to be to have a measurable effect on the Earth, and what evidence might be left behind in the geological record? \citet{1977Natur.265..318C} considered a scenario in which roughly one supernovae occurs within 10 pc of Earth every time we cross a spiral arm of the Milky Way. As argued by \citet{1974Sci...184.1079R} and \citet{1976Natur.263..398W}, while radiation would be absorbed in the atmosphere and so not reach Earth's surface, the ozone layer would be destroyed for some period of time, perhaps years. As a result, harmful solar ultraviolet radiation could destroy significant amounts of surface marine life, collapsing food chains and leading to a mass extinction. Also, cosmic ray flux will increase, and for supernovae within $\sm 20$ pc, the ejecta shell may overcome the solar wind ram pressure; both events leading to an increased flux of of DNA-damaging radiation reaching the Earth's surface \citep{2017hsn..book.2419K}. More detailed calculations of this scenario have been presented by \citet{1995PNAS...92..235E}, \citet{1996PNAS...93.1582C}, \citet{2003ApJ...585.1169G}, \citet{2011Ap&SS.336..287B}, \citet{2012MNRAS.423.1234S}, \citet{2012IJAsB..11..279H} \citet{Melott2018}. In particular, \citet{Melott2017} argue that muons will irradiate the ocean to a depth of a kilometre. Being more penetrating than other natural forms of radiation, muons would particularly affect ocean megafauna.

Similarly, \citet{1995ApJ...444L..53T} argue that gamma ray bursts would have a similar effect, though at larger distances. \citet{2005ApJ...622L.153T,2005ApJ...634..509T} model the effect of a GRB at a ``typical'' distance of 2 kpc, showing that about a third of the ozone is depleted for several years, leading to 16 times greater DNA damage and a cooling of the climate. Such an event is a candidate for the cause of the late Ordovician mass extinction. Using an analytic model of the Milky Way, \citet{2014PhRvL.113w1102P} calculate that there is a 50\% of an ozone-removing, extinction-causing GRB in the vicinity of Earth in the last 500 Myr.

More extreme sources of energy will cause more lasting damage. \citet{2017NatSR...7.5419S} note that $5.6 \ten{33}$ ergs is sufficient to boil Earth's ocean, that is, raising its temperature by more than 100$^\circ$C. A supernova of energy $10^{51}$ erg would need to explode within 0.04 pc for enough energy to be incident on Earth for this to happen, while a GRB that blasts a similar amount of energy into a $\sm 2^\circ$ beam could be as far as $\sm$13.8 pc. On longer timescales, high-energy photons from a quasar or active galactic nucleus (AGN) can strip the atmosphere and even the ocean from a planet. This effect can accumulate over many millions of years, as the atmosphere or ocean is gradually unbound from Earth and lost into space. 

The effect of cosmic events and environments has been applied to the GHZ. \citet{2004Sci...303...59L} characterised the GHZ of our galaxy, taking into account the effect of supernovae close to the galactic centre, the distribution of metals, and the delay between star formation and the formation of life on a planet. They find a relatively narrow zone, located away from the galactic centre. \citet{2008SSRv..135..313P} and \citet{2011AsBio..11..855G} refine and extend the calculation of the GHZ, noting some significant sources of uncertainty. \citet{2017IJAsB..16...60F} use high-resolution simulations of the Local Group of galaxies to model the habitable zone of the Milky Way and M33, taking into account planet formation and giant planet migration, as well as the probability of nearby supernovae. They find generally broad GHZs, extending from the centre of the galaxy to $\sm 10$ kpc for most of cosmic history.

On cosmic scales, \citet{2016arXiv160609224D} use an analytical model of the star-formation history of the universe to calculate the relative effect of cosmic explosions and quasars on life. They conclude that Type II (core-collapse) SN and Type Ia SN dominate the production of life-affecting radiation, with negligible contribution from GRBs and quasars. \citet{2018MNRAS.475.1829S} explore the GHZ using the Millennium N-body simulation and a semi-analytic model of galaxy formation. This allows them to take into account the metallicity dependence of planet formation and extinction events (SN, GRBs and AGN). They find that 18\% of stars today have been ``irradiated'' by a high-energy cosmic event in the last 260 Myr, and that this percentage was much higher in the past.

Some attention has also been given to cosmic \emph{time} scales, following the habitability of the cosmos over time. \citet{2016JCAP...08..040L} argue that a life-form is most likely to observe that they are on a planet orbiting a $\sm 0.1$ \Msol star, ten trillion years after the big bang.

In a previous paper \citep{2018MNRAS.477.3727B}, we considered cosmic habitability across different values of the cosmological constant using the \eagle galaxy formation code \citep{2015MNRAS.446..521S}. We did not consider extinction events, assuming that the production of observers is proportional to the production of planets, which depends on the formation of stars and the production of elements in stars. However, such events have been argued to be potentially very significant, particularly in universes with different fundamental parameters. \citet[][see also \citealt{2006PhRvD..73b3505T,2015JCAP...09..030A}]{1998ApJ...499..526T} considered the effect of star-star interactions, wherein a nearby star could perturb a planet into a highly eccentric orbit, or even throw it clear into interstellar space. These effects become particularly significant as the amplitude of primordial density fluctuations ($Q$) increases. Supernovae, both core collapse (CC) and Type Ia (SNIa) will also cause extinction events: \citet{2019AN....340..173T} argue that taking into account the probability of SN in regions of high stellar density substantially changes the expected value of $\Lambda$ in multiverse models, to be more in line with observations. Gamma ray bursts (GRBs), too, will extinguish life within a certain surrounding region. \citet{2016PhRvL.116h1301P} argue that because GRBs occur in low metallicity ($Z < 0.3$) environments and have a large ``extinction'' radii ($\sm 20 - 50$ kpc), habitable galaxies need to be free from satellite galaxies (and their low-metallicity, GRB-producing gas and stars) in their inner regions. For moderate values of $\Lambda$ (from zero to a few times $\Lambda_0$), the universe produces more ``isolated'', habitable haloes as $\Lambda$ increases. Perhaps counterintuitively, this implies that a (positive) cosmological constant that is \emph{smaller} than that in our universe can be detrimental to life.

In this paper, we will apply a more detailed model of observer production and survival to the \eagle galaxy formation simulations, which includes the effect of cosmic extinction processes: core collapse SN, Type Ia SN, star-star interactions, GRBs and quasars. In Section \ref{S:sims}, we briefly review the \eagle galaxy formation code and simulations. In Section \ref{S:obs}, we describe our model for the effect of extinction processes on the habitability of planets.

\section{Galaxy Formation Simulations} \label{S:sims}
The Virgo Consortium's \textsc{\eagle} project (\emph{Evolution and Assembly of GaLaxies and their Environment}) is a suite of hydrodynamical simulations that follow the formation of galaxies and supermassive black holes in cosmologically representative volumes of a standard $\Lambda$CDM universe. The details of the code, and particularly the sub-grid models, are described in \citet{2015MNRAS.446..521S}. The initial conditions for the \eagle simulations were set up using a transfer function generated using CAMB \citep{2000ApJ...538..473L} and a power-law primordial power spectrum with index $n_s = 0.9611$. Particles were arranged in a glass-like initial configuration, and were displaced according to second-order Lagrangian perturbation theory \citep{2010MNRAS.403.1859J}. The code models the effect of radiative cooling for 11 elements, star formation, stellar mass loss, energy feedback from star formation, gas accretion onto and mergers of supermassive black holes (BHs), and AGN feedback. Black holes are seeded in all dark matter haloes with masses greater than $10^{10} h^{-1} \Msol= 1.48 \ten{10} \Msol$.

The halo finding algorithm is described in \citet{2015MNRAS.446..521S}; in short, the code regularly runs the friends-of-friends (FoF) finder \citep{1985ApJ...292..371D} with linking length 0.2 on the dark matter distribution. When \emph{calculating} the habitable zone in the following sections, we use membership in the largest halo to identify the widest relevant neighborhood of a given star particle. Thus, we use the FOF algorithm to identify haloes. When we later come to discuss ``galaxies'' and their occupants, we follow the usual approach of the \textsc{\eagle} project and use the \emph{smallest} bound substructure to identify galaxies.

Here, we analyse the ``Reference'' $\Lambda$CDM universe \eagle simulation, which uses the \citet{2014A&A...571A..16P} cosmological parameters: $(\Omega_m,\Omega_\Lambda,\Omega_b,h,\sigma_8,n_s,Y) = (0.307,0.693,0.04825,0.6777,0.8288,0.9611,0.248)$.  The box is 100 comoving-Mpc in size, and contains $1504^3$ dark matter particles, each with mass of $9.7 \ten{6}\Msol$. Initially, the simulation contains an equal number of baryon particles, each with mass of $1.81 \ten{6}\Msol$. The gravitational softening length is 2.66~kpc comoving or 0.70~kpc proper, whichever is smaller.

The \eagle simulations are known to reproduce a wide range of observations of the low redshift universe. \citet{2015MNRAS.446..521S} note that the galaxy stellar mass function, $M_*$-$M_{200}$, galaxy sizes and $M_\ro{BH}$-$M_*$ relation were used to callibrate various subgrid parameters. Other relationships that match with observations are successful predictions of the model. No model is perfect, however, and there are known discrepancies regarding the metallicities of dwarf galaxies and the intracluster medium. Further, the simulation box size determines the largest structures that form. For example, the population of galaxies recently discovered at $z > 8$ by JWST are too large (at that time) to be present in the \eagle simulation box. As a result, the simulations will not capture the population of large galaxies in our universe, particularly ellipticals. We will return to these limitations in our conclusions.

\section{Computing The Galactic Habitable Zone} \label{S:obs}

A given snapshot of an \textsc{\eagle} simulation contains different types of particles. Throughout, we will use the subscript `$i$' to refer to the star particle (really, a population of stars) whose habitability we are considering, `$j$' to index neighbouring star particles, and `$k$' to index neighbouring gas particles. We will not need to consider dark matter particles.

\paragraph*{Star particles} represent a stellar population,
\begin{itemize}
\item centred at position $\bm{x}_i$,
\item total stellar mass $m^*_i$,
\item metallicity $Z_i$, and 
\item born at time $t_i$.
\end{itemize}

\paragraph*{Gas (or SPH) particles} represent interstellar/intergalactic gas,
\begin{itemize}
\item centred at position $\bm{x}_k$,
\item total baryonic mass $m^g_k$,
\item smoothing length $h_k$,
\item metallicity $Z_k$, and 
\item star formation rate $\dot{N}^*_k$.
\end{itemize}
 We use the simulation snapshot to calculate the expected number of life-forms that exist in the simulation volume at each time. The major unknown in our calculation is the probability that life will form per unit time on a suitably habitable planet. Given the uncertainties of abiogenesis, we do not wish to guess this probability. However, we will assume that, having taken into account the average conditions that produce habitable planets, there is a fixed probability ($f_l$) that life will form on a habitable planet per unit time.

We are especially interested here in the effect of astronomical processes that make an otherwise suitable planet unsuitable for life. We distinguish between two grades of catastrophe. Firstly, there are astronomical events that render a planet \emph{uninhabited}, that is, which kill all existing life on a planet. For example, it is highly unlikely that life on Earth could have survived the impact of the hypothesised Mars-sized object Theia that lead to the formation of our Moon. But this impact (assuming it occurred) obviously did not render Earth uninhabitable. An event that renders a planet uninhabited we will call an \emph{extinction}.

Secondly, there are more extreme astronomical events that render a planet \emph{uninhabitable}. They not only kill any existing life but make the future (re-)formation of life effectively impossible. For example, if a cosmic catastrophe were to remove the Earth's atmosphere and oceans today, and internal and external processes are judged unlikely to replenish the supply of water and oxygen, then we would expect the Earth to remain thereafter uninhabitable. An event that renders a planet uninhabitable we will call a \emph{razing}.\footnote{Our referee suggests that we could model both kinds of events by rendering the planet uninhabited for time $t_u$. Then, extinction would be modelled as $t_u = 0$, and razing as $t_u \rightarrow \infty$. However, in the absence of a model for $t_u$ as a function of distance etc. for each catastrophe type, we leave this for future work.}

For each of our cosmic explosions (Core-Collapse SN, Type Ia SN and GRBs), we will define an \emph{extinction} radius and a smaller \emph{razing} radius. A star within the extinction radius will be rendered uninhabited. A star within the razing radius will be rendered uninhabitable.

Given our assumptions, the expected number of observers that exist in the simulation box at a given time is calculated by summing over all the star particles,
\begin{equation} \label{eq:Nobs}
N_\ro{obs}(t) = f_l \sum\limits_i \frac{m^*_i}{\bar{m}_*} ~ f_\ro{ms} (t - t_i) ~ P_\ro{HP}(Z_i) ~ T_{H,i}(t) ~ P_\ro{nr}(t_i,t) ~ .
\end{equation}
The terms in the equation are as follows.

\paragraph*{Number of stars:} $m^*_i / \bar{m}_*$ is the number of stars in the star particle, where $\bar{m}_*$ is the average stellar mass. Using the Chabrier IMF, with $M_{min} = 0.1 \Msol$ and $M_{max} = 100 \Msol$ as in the simulation's physics, the average stellar mass is $\bar{m}_* = 0.795 \Msol$.

\paragraph*{Main sequence fraction:} $f_\ro{ms}(\Delta t)$ is the fraction of stars (by number) in the population that are still on the main sequence a time $\Delta t$ after they are born. As in B18, we calculate $f_\ro{ms}$ using the \citet{2003PASP..115..763C} initial mass function, and a simple relationship between mass and main-sequence lifetime drawn from the analytic model of \citet{2008JCAP...08..010A} normalised to $t_\ro{ms} = 10$ Gyr at Solar mass; this broadly consistent with \citet{1998A&A...334..505P}.

\paragraph*{Probability of a habitable planet:} $P_\ro{HP}(Z_i)$ is the probability that a main-sequence star of given metallicity $Z_i$ hosts a terrestrial planet in the stellar habitable zone, in a system in which a close-in giant star will/has not formed. Studies of exoplanets have found a large population of ``hot jupiters'' or close-in giant planets \citep{2025PSJ.....6..186C}. Because these are unlikely to form in-situ, they mostly form at larger distances and migrate inwards, engulfing or unbinding any planet in the habitable zone. To permit easier comparison with previous calculations, we use the prescription of \citet{2001Icar..151..307L}. The probability of an Earth-like planet $(P_\ro{EP})$ rises from zero at $Z = 0.1 \Zsol$ to unity at $\log(Z/\Zsol) = 0.6$. The probability of a close-in giant planet $(P_\ro{GP})$ rises quickly from zero at $\log(Z/\Zsol) = -0.1$ to unity at $\log(Z/\Zsol) = 0.4$. Then,
\begin{equation}
P_\ro{HP}(Z_i) = P_\ro{EP}(Z_i) ~~ [ 1 - P_\ro{GP}(Z_i) ] ~.
\end{equation}
We use a fit to the curve plotted in Figure 1 of \citet{2004Sci...303...59L}.

\paragraph*{Habitable time:} $t_{H,i}(t)$ is the average amount of time that an average planet around a star $i$ has been habitable at time $t$, since either the star was born or the last (total) extinction event. Given a Poisson distribution of extinction events with average rate $r_e(t)$ in the neighbourhood of a given star, $P_{ne}(t_1,t_2)$ is the probability of no extinction events between $t_1$ and $t_2$,
\begin{equation} \label{eq:Pne}
P_{ne}(t_1,t_2) = \exp \left(-\int_{t_1}^{t_2} r_e(t') \df t' \right).
\end{equation}
Then, the expectation value of the time since the most recent extinction event is,\footnote{For calculating this quantity from successive snapshots, we use an equivalent ordinary differential equation: $\dd T_{H,i} / \dd t = 1 - r_e(t)~T_{H,i}$.}
\begin{equation} \label{eq:tHi}
T_{H,i}(t) = \int_{t_i}^{t} P_{ne}(t_e,t) ~ \df t_e ~.
\end{equation}

\paragraph*{Probability of no razing:} $P_\ro{nr}(t_i,t)$ is the probability of no razing events, which render the planet uninhabitable, occurring between $t_i$ and $t$. Similarly to extinction, assuming a Poisson process and an average event rate $r_r(t)$,
\begin{equation} \label{eq:Pnr}
P_\ro{nr}(t_i,t) = \exp \left(-\int_{t_i}^{t} r_r(t') \df t' \right) ~.
\end{equation}

\begin{table}
\begin{center}
\begin{tabular}{|l|ll|}
\hline
Parameter		& 				 & Fiducial Value \\
\hline 
$\bar{m}_*$ 	& average mass per star							& 0.795 \Msol\\
$d_\ro{Ce}$		& CCSN extinction radius						& 0.3 pc\\
$d_\ro{Cr}$		& CCSN razing radius								& 0.01 pc\\
						& CCSN minimum stellar mass				& 6 \Msol\\
$f_\ro{CC}$			& CCSN per star										& 0.014\\
$d_\ro{Ia,e}$	& Type Ia SN extinction radius					& 0.3 pc\\
$d_\ro{Ia,r}$	& Type Ia SN razing radius						& 0.01 pc\\
$f_\ro{Ia}$ 		& Type Ia SN per star 								& 1.6 \ten{-3}\\
$\tau$ 				& Type Ia SN timescale								& 2 Gyr\\
$n$ 					& $n^{th}$ nearest stellar nbr 				 & 5\\
$\sigma_0$		& planetary system cross section 			& 1 AU$^2$\\
$\theta_b$ ($f_b$)	& GRB jet half-radius (beaming) 		& $2^{\circ}$ (1/1600)\\
						& observed global GRB rate						& $10^9$ GRB/Gpc$^3$/Gyr \\
$d_\ro{Ge}$		& GRB extinction radius							& 12 pc\\
$d_\ro{Gr}$		& GRB razing radius									& 0.4 pc\\
$\Phi_X$			& quasar razing fluence							& $2.5 \ten{16}$ erg/cm$^2$\\
\hline
\end{tabular}
\caption{Model parameters for extinction and razing, as described in Section \ref{S:obs}.} \label{tab:params}
\end{center}
\end{table}

\subsection{Extinction and Razing Events}
As mentioned, we consider five types of astronomical events that contribute to the total extinction and razing rate: CCSN, SNIa, GRBs, star-star interactions, and quasars (QSOs). Models of the effect of cosmic explosions on our planetary environment focus on the crucial quantity \emph{fluence} ($\Phi$), the incident energy per unit area,
\begin{equation} \label{eq:rEphi}
\Phi = \frac{E}{\Omega ~ r^2} ~,
\end{equation}
where $r$ is the distance from the planet to the explosion or quasar, and $E$ is the energy of life-damaging radiation which is emitted into solid angle $\Omega$; if the emission is isotropic, $\Omega = 4\pi$. If fluence $\Phi$ is absorbed by the cross-section of Earth, the energy absorbed is,
\begin{equation}
E_\ro{abs} = 1.3 \ten{26} ~\ro{erg} ~ \frac{\Phi}{10^8 \ergcm} ~.
\end{equation}
By way of comparison, the world's entire nuclear arsenal is $\sim$1,500 megatons of TNT, which is $6 \ten{25}$ erg.

The damage done to life depends on the fluence at Earth. On the lower end, according to \citet{1995ApJ...444L..53T} and \citet{2005ApJ...634..509T}, a fluence of $10^8 \ergcm$ from a GRB would destroy a substantial fraction of the ozone layer, resulting in an enhancement of ultraviolet solar radiation that would kill surface marine life, greatly increase DNA damage, and possibly causing global cooling and nitric acid rain. If such an event is indeed responsible for the Late Ordovician mass extinction, then the aftermath caused a loss of 85\% of marine species. However, life on Earth returned to previous levels of biodiversity after $\sm 10$ million years, which is a minor delay on the timescale of a typical main-sequence star and its planets. This event obviously did not preclude the later development of intelligent life on Earth. We are focussed here on astrophysical events with more extreme, more permanent consequences.

At the most extreme end, \citet{2018MNRAS.479..171F} calculate the total surface mass loss (atmosphere and/or ocean) from a planet that has absorbed fluence $\Phi_X$ from a nearby quasar,
\begin{equation}
M_\ro{lost} = 3 \epsilon ~ (4 \pi G \rho_P)^{-1} ~ \Phi_X ~,
\end{equation}
where $\Phi_X = \int F_X \df t$ is specifically the extreme UV (XUV, with energy $> 10$ eV) fluence, $F_X$ is the XUV flux, $\epsilon$ is the fraction of the incident power available to heat the atmosphere (with fiducial value 0.1), and $\rho_P$ is the average density of a rocky planet. The restriction to XUV radiation makes the timescale of absorption irrelevant, as each photon has roughly enough energy to give an atom in the atmosphere/ocean escape velocity. Using typical values, \citeauthor{2018MNRAS.479..171F} calculate that a fluence of $\Phi_X = 2.5 \ten{16}$ erg/cm$^2$ would be sufficient to remove (mass equivalent to) the Earth's atmosphere, and $6.3 \ten{18}$ erg/cm$^2$ will remove the Earth's ocean.

Recall the calculation of \citet{2017NatSR...7.5419S} above: $5.6 \ten{33}$ erg of energy to boil Earth's oceans. This implies a fluence of $4.4 \ten{15}$ erg/cm$^2$, given Earth's cross-sectional area.

These are simple energy arguments; most of the detailed modelling has investigated the more modest scenario where only 85\% of life on Earth goes extinct. In addition to the required energy, we need to consider corrections that account for the fraction of the total energy output that is in a lethal form. In light of this, we apply order-of-magnitude limits of $\Phi_e \approx 10^{14}$ erg/cm$^2$ for extinction and $\Phi_r \approx 10^{17}$ erg/cm$^2$ for razing, for the total energy absorbed by Earth. We will now consider each astrophysical event or process in turn.

\paragraph*{Core-Collapse supernovae (CCSN) :} Let $\pcc(t) \df t$ be the probability that a randomly chosen star goes CCSN between $t$ and $t + \df t$ after its birth. The normalisation $f_\ro{CC} = \int_0^\infty \pcc(t) \df t$ is the fraction of all stars formed (by number) that go CCSN at some point in their life.

Given that the scale of the SN explosion is not resolved in simulation snapshot, we model the effect of CCSN on life as a single extinction/razing ``blast radius'': any planet within $d_\ro{Ce}$ ($d_\ro{Cr}$) of the explosion --- corresponding to a volume of $V_\ro{Ce} = 4/3 \pi ~ d_\ro{Ce}^3$ ($V_\ro{Cr}$) --- will be rendered uninhabited (uninhabitable). Consider first the case of an extinction event. For a given star that potentially hosts life, let $P_e(t_e,t_f) \df t_e \df t_f$ be the probability that an extinction event occurs between $t_e$ and $t_e + \df t_e$, within a volume $V_\ro{Ce}$ around the star, due to a neighbouring star that formed between $t_f$ and $t_j + \df t_f$,
\begin{equation}
P_e(t_e, t_f) \df t_e \df t_f = V_\ro{Ce} ~ \dot{n}_*(\bm{x},t_f) ~ \pcc(t_e - t_f) ~ \df t_e \df t_f ~,
\end{equation}
where $\dot{n}_*(\bm{x},t_f)$ is the local star-formation rate (by number) as a function of position ($\bm{x}$) and time. (Strictly speaking, we should be integrating over the volume $V_\ro{Ce}$, but the simulations do not predict variations of the star formation rate on these small scales.) Then, the average rate of extinction events for a given star is calculated by integrating over all the neighbouring stars that formed before the present time $t$,
\begin{equation} \label{eq:rCe}
r_{Ce}(t) = V_\ro{Ce} ~ \int_0^{t} \dot{n}_*(\bm{x},t_f) ~ \pcc(t - t_f) \df t_f ~.
\end{equation}

CCSN occur when stars with mass greater than 6-8 \Msol reach the end of stellar burning \citep{2009ARA&A..47...63S}, which is less than 40 million years after their birth \citep{2012MNRAS.427..127B,2012MNRAS.423.1234S,2013EPJWC..4303001B}. This is short enough, on the timescale of our simulations, to treat CCSN as instantaneous, so that the probability is a Dirac-delta function of time: $\pcc(t) \approx f_\ro{CC}~ \delta (t)$. Then, for $t > t_i$,
\begin{align}\label{eq:reCC}
r_{Ce} (t) &\approx V_\ro{Ce} ~ f_\ro{CC} ~ \dot{n}_*(\bm{x}_i,t) ~ \\
&\approx 0.04~\ro{Gyr}^{-1} \left(\frac{V_\ro{Ce}}{0.1~\ro{pc}^3}\right) \left(\frac{f_\ro{CC}}{0.014}\right) \left(\frac{\dot{n}_*}{24~\ro{stars/yr/kpc}^3}\right)
\end{align} 
where $\dot{n}_*(\bm{x}_i,t)$ is the star formation rate (by number) at $t$ that formed in the volume $V_\ro{Ce}$. The second equation uses fiducial values from Table \ref{tab:params}, and star forming properties of the starburst galaxy M82: 10 $\Msol$/yr in the central nucleus of radius 0.5 kpc. For a Milky-Way like galaxy, with 1 $\Msol$/yr of star formation in a large disk, $r_{Ce} \sim 10^{-5}~\ro{Gyr}^{-1}$.

We calculate the local star formation rate at all times that the star particle exists, not just when it was born. The term $r_{Ce}$ is added to the total extinction rate $r_e$ in Equation \eqref{eq:Pne}. The same formalism applies to razing: by replacing $V_{Ce}$ with $V_{Cr}$ in Equation \eqref{eq:reCC}, we calculate the contribution of CCSN to the total razing rate $r_r$.

We calculate $\dot{n}_*$ in our simulations using a sum over neighbouring SPH particles. In the simulation output, each gas particle has an associated star formation rate by mass (SFR$_k$), which we can relate to the star formation rate by number ($\dot{N}^*_k$) using the average stellar mass $\bar{m}_*$. The star formation \emph{density} (by number) at the position ($\bm{x}_i$) of the star particle is,
\begin{equation}
\dot{n}_*(\bm{x}_i,t) = \sum\limits_k \dot{N}^*_k ~ W(\bm{x}_i - \bm{x}_k;h_k) ~,
\end{equation}
where $W$ is the SPH smoothing kernel, and $h_k$ is the SPH smoothing length. Finally, assuming that all stars with masses greater than 6 \Msol explode as CCSN, we use the Chabrier IMF to calculate the CCSN fraction per star, $f_\ro{CC} = 0.014$. 

While CCSN exhibit a range of explosion energies, given the approximations made in assigning extinction/razing fluences, we assume a single fiducial isotropic CCSN energy of $10^{51}$ erg \citep{1998PASP..110..125J}. With fluences of $\Phi_e \approx 10^{14}$ erg/cm$^2$ for extinction and $\Phi_e \approx 10^{17}$ erg/cm$^2$ for razing, Equation \eqref{eq:rEphi} tells us that a CCSN will cause extinction at $d_{Ce} = 0.3$ pc and razing at $d_{Cr} = 0.01$ pc.

\paragraph*{Type Ia supernovae (SNIa):} We can use much of the formalism developed above. A given star will explode as a Type Ia SN if it ends its life as a white dwarf in a binary system, and accretes mass from a companion until the Chandresekar limit is crossed. An empirical relation for the rate of SNIa per unit stellar mass is incorporated into the \eagle feedback physics,
\begin{equation}
\dot{N}_\ro{Ia} = \nu \frac{\exp(-t / \tau)}{\tau} ~,
\end{equation}
where $\nu = 2 \ten{-3}$ $\Msol^{-1}$ and $\tau = 2$ Gyr. We can use the the average stellar mass to convert to the SNIa rate per star: $\pia = \bar{m} \dot{N}_\ro{Ia}$, so that $f_\ro{Ia} = \bar{m} \nu = 1.6 \ten{-3}$ Type Ia SN per star formed.

As above, we consider first the case of extinction. For a potentially life-hosting star born at $t_i$, we consider the contribution to its neighbourhood from another star particles in the simulation, born at time $t_j$ and with associated stellar density $n^*_j$. We can treat its contribution to the star formation rate in Equation \eqref{eq:rCe} as a Dirac delta function\footnote{This is not equivalent to setting the supernovae probability to be a delta function, as we did with \pcc above.} ($\dot{n}^*_j(t) = n^*_j \delta(t-t_j)$),
\begin{align*}
r_{Ia,e} (t;t_j) &= V_\ro{Ia,e} ~ f_{Ia} ~ n^*_j ~ \frac{\exp(-(t-t_j) / \tau)}{\tau} \\
&\approx 1.3 \ten{-4}~\ro{Gyr}^{-1} \left(\frac{V_\ro{Ia,e}}{0.1~\ro{pc}^3}\right) \left(\frac{f_\ro{Ia}}{0.014}\right) \\
&\qquad \times \left(\frac{n^*_j}{1.4~\ro{stars/pc}^3}\right) \left( \frac{E_\tau}{(2~\ro{Gyr})^{-1}} \right)
\end{align*} 
where $V_\ro{Ia,e}$ is the extinction volume of a SNIa. The second equation uses fiducial values from Table \ref{tab:params}, $E_\tau$ is the exponential term in the line above, and stellar properties of the starburst galaxy M82: $6 \ten{8}~\Msol$ in the central nucleus of radius 0.5 kpc. For a Milky-Way like galaxy, with $2\ten{11}$ stars spread over the thin disk, $r_{Ia,e}$ is similar.

%

Because of the Gyr timescales between stellar birth and SNIa, we need to take into account changes in the local stellar density over the lifetime of the star particle. Star particles are born in regions of high gas density, and so the initial environment of an individual star depends on scales smaller than the population of stars represented by simulation star particle. However, on timescales relevant to life and SNIa, we expect stars to have mixed into their wider galactic environment. As a result, we treat our star particles as tracers of the overall stellar density. Any subgrid effect of stellar clustering is be folded into the overall dependence on density. 

We calculate the stellar density by summing over neighbouring star particles. Because star particles don't have an associated smoothing length, we use an $n$-th nearest neighbour approach: for particle $i$, we find the distance $d_n$ that encompasses the $n$ closest neighbours. Then, the stellar number density is $n_{*} = (4/3 \pi ~ d_n^3)^{-1} ~\sum_j m_j$, summing over the $n$ neighbouring particles $j$ and the particle $i$.

We calculate $r_{Ia,e}$ by averaging over the $n$ neighbouring particles and the particle $i$, weighting each contribution by the SPH kernel function $W$, with smoothing length given by $d_n$,
\begin{equation}
r_{Ia,e} (t) = V_\ro{Ia,e} ~ n^*_{i} ~ \frac{f_{Ia}}{\tau} ~ \frac{\sum_{j} m^*_j ~ W(d_n) \exp(-(t-t_j) / \tau) }{\sum_{j} m^*_j ~ W(d_n) } ~.
\end{equation}
As before, the same formalism applies to razing. We place the razing volume $V_\ro{Ia,r}$ in the equation above, and it gives us the Type Ia contribution $r_{Ia,e}$ to the total razing rate.

As with CCSN, we assume a single fiducial isotropic SNIa explosion energy of $10^{51}$ ergs \citep{1998PASP..110..125J}. Then, Equation \eqref{eq:rEphi} tells us that a SNIa will cause extinction at $d_{Ia,e} = 0.3$ pc and razing at $d_{Ia,r} = 0.01$ pc.

\begin{figure} \centering
	\includegraphics[width=0.45\textwidth]{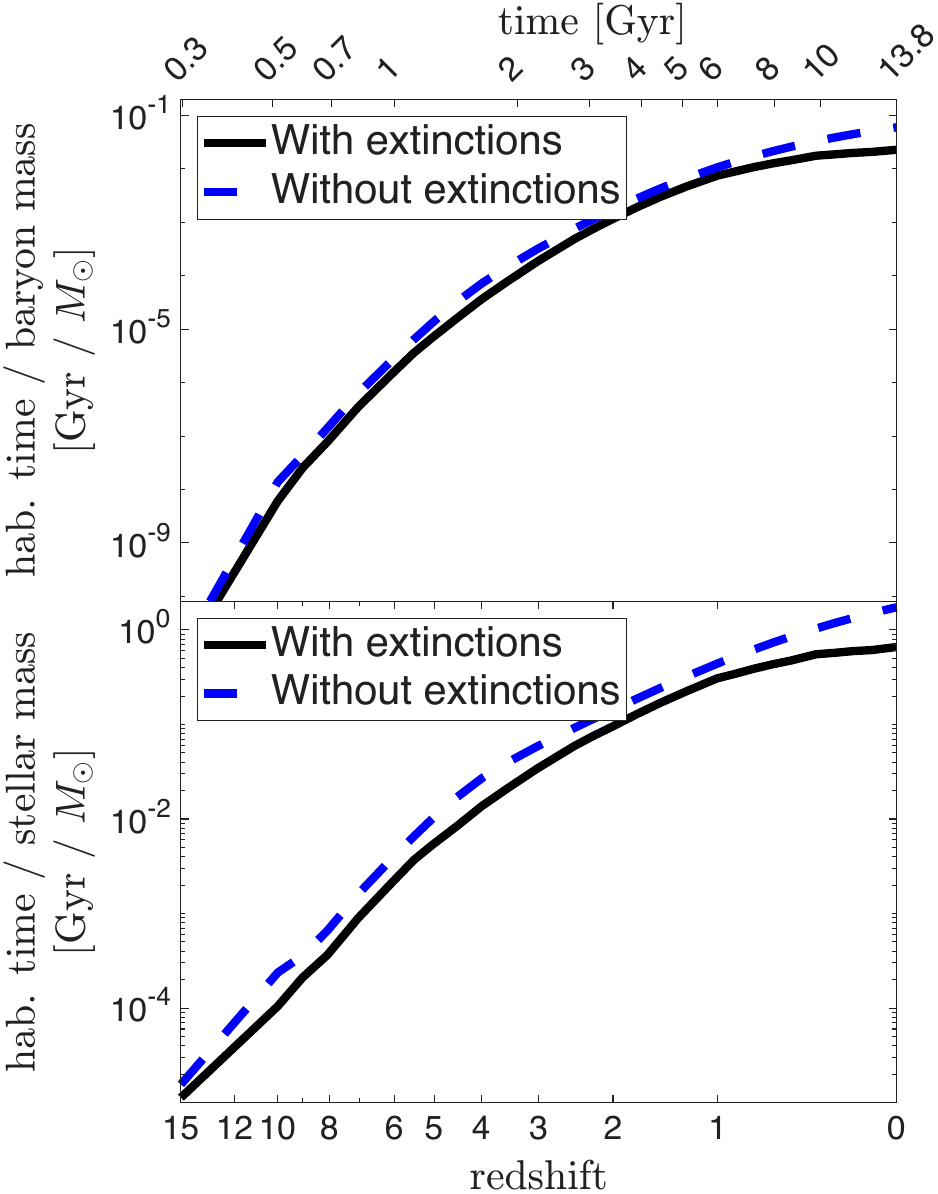}
	\caption{The habitable time per unit baryonic mass (top) and per stellar mass (bottom), as a function of cosmic time (top axis) and redshift (bottom axis). Dashed lines show the predictions with no extinction events, while the solid line shows predictions will all extinctions. The plotted quantities (habitable time) can be interpreted as follows: a value of $1 \ten{-3} \ro{Gyr} / \Msol$ in the top (bottom) plot means that, on average, each solar mass of baryonic (stellar) matter at time $t$ has produced a million years of life-friendly time on suitable planets. Both curves increase rapidly as early star formation creates suitable stars, metals, and planets. The ``extinctions'' (solid) line flattens at late times.} \label{fig:Nobs_vs_t}
\end{figure}

\begin{figure} \centering
	\includegraphics[width=0.45\textwidth]{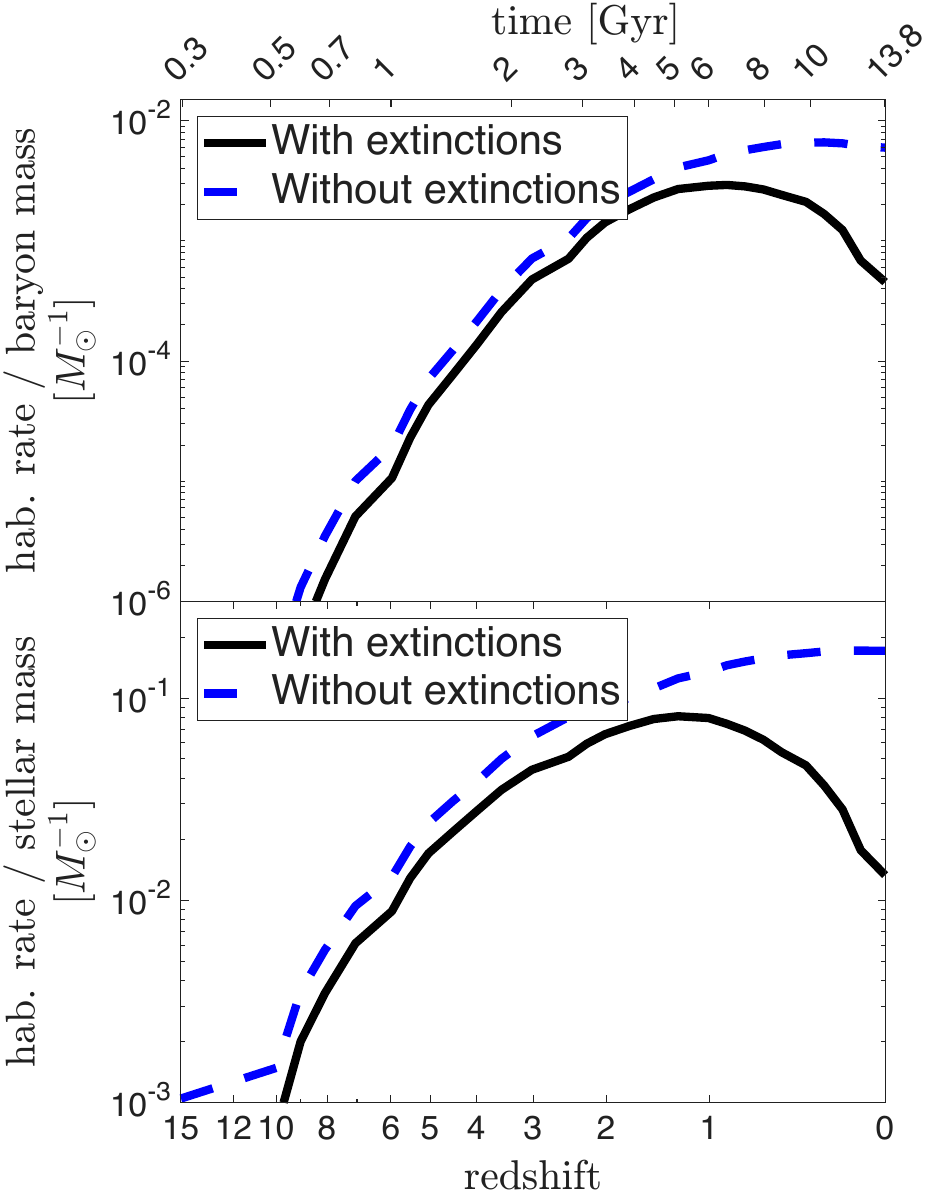}
	\caption{The habitable time \emph{rate} per unit baryonic mass (top) and per stellar mass (bottom), calculated with (solid) and without (dashed) extinction events. This is the time derivative of Figure \ref{fig:Nobs_vs_t}.The plotted quantities (habitable time rate) can be interpreted as follows: a value of $1 \ten{-3} \Msol^{-1}$ in the top (bottom) plot means that, on average, for every billion years that passes and for each solar mass of baryonic (stellar) matter, a million years of life-friendly time passes on suitable planets. Both curves increase rapidly as with star formation, but turn over in the late universe, as extinction renders planets uninhabited/uninhabitable faster than they can be made.} \label{fig:dNobs_vs_t}
\end{figure}

\begin{figure*} \centering
	\begin{minipage}{0.48\textwidth}
		\includegraphics[width=\textwidth]{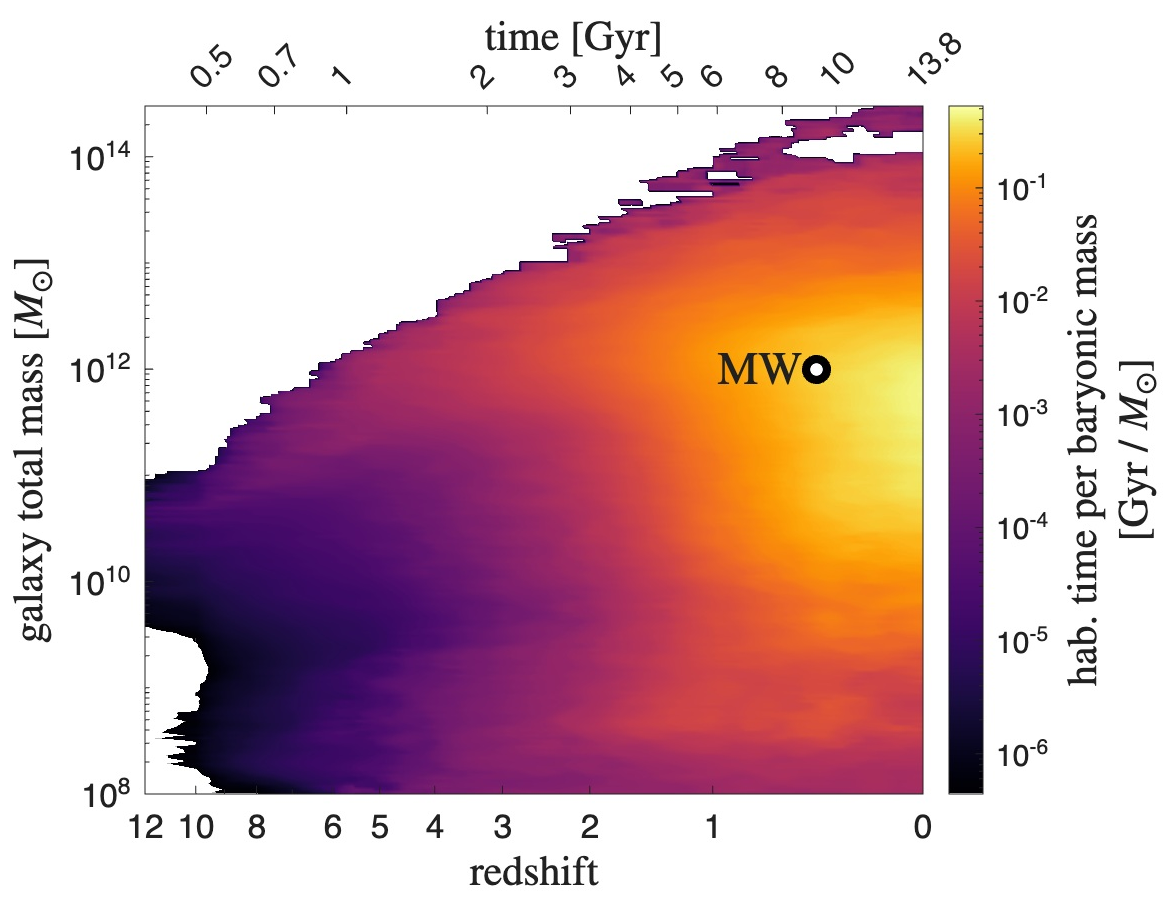}
	\end{minipage}
	\begin{minipage}{0.48\textwidth}
		\includegraphics[width=\textwidth]{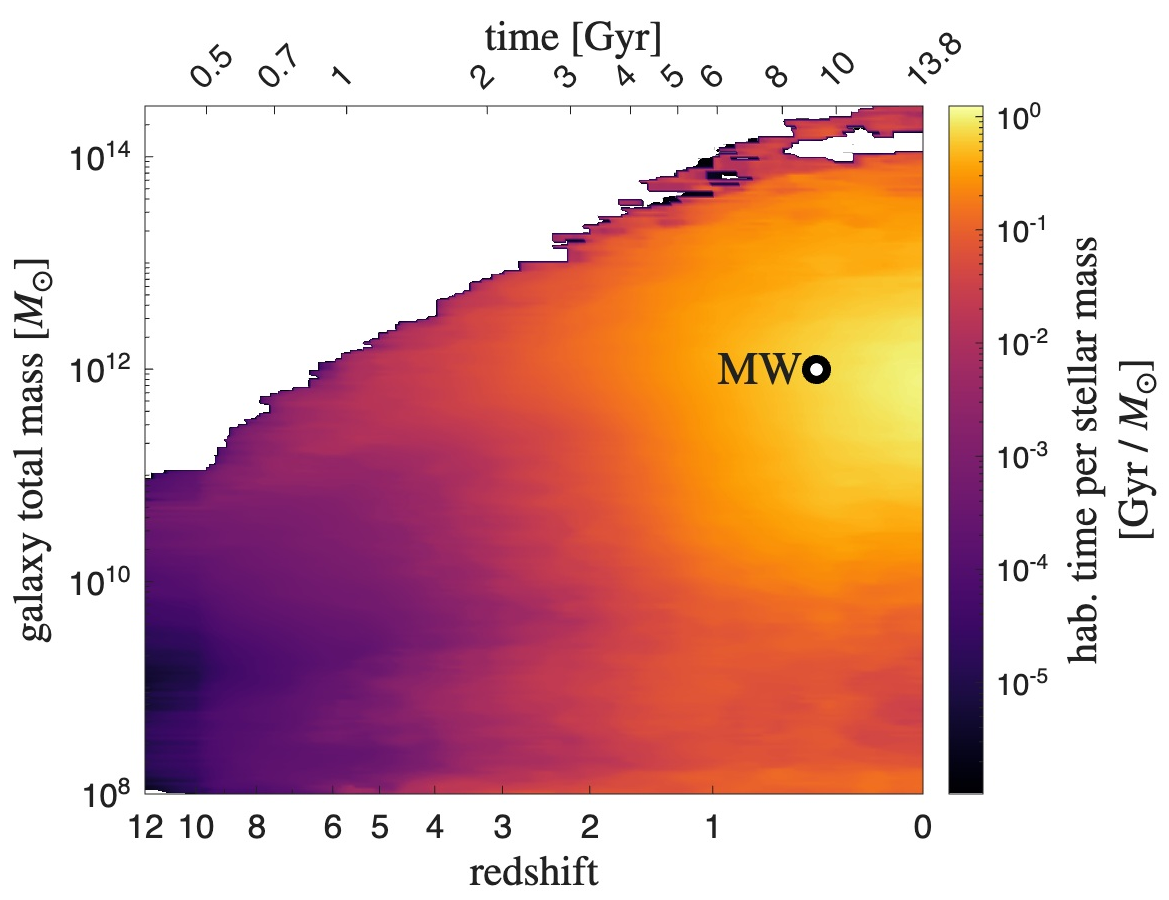}
	\end{minipage}
	\begin{minipage}{0.48\textwidth}
		\includegraphics[width=\textwidth]{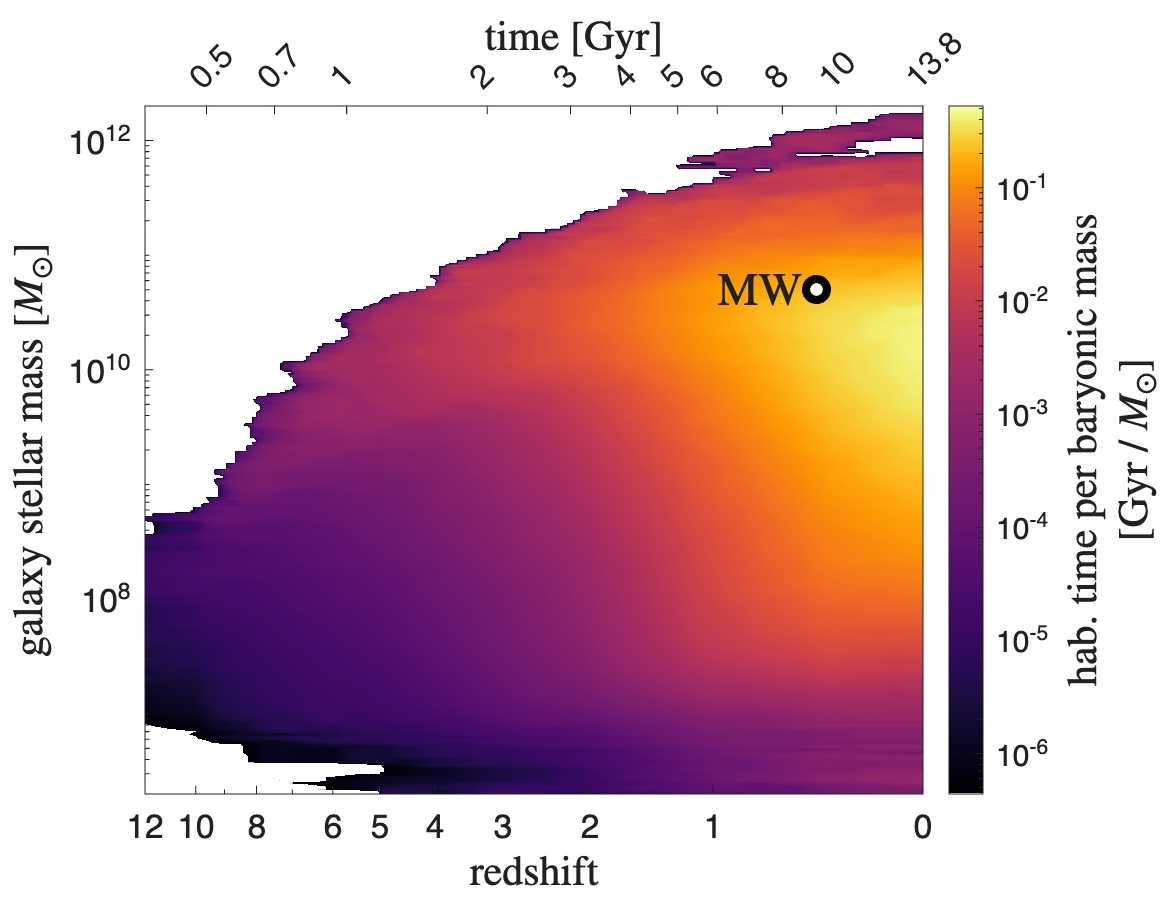}
	\end{minipage}
	\begin{minipage}{0.48\textwidth}
		\includegraphics[width=\textwidth]{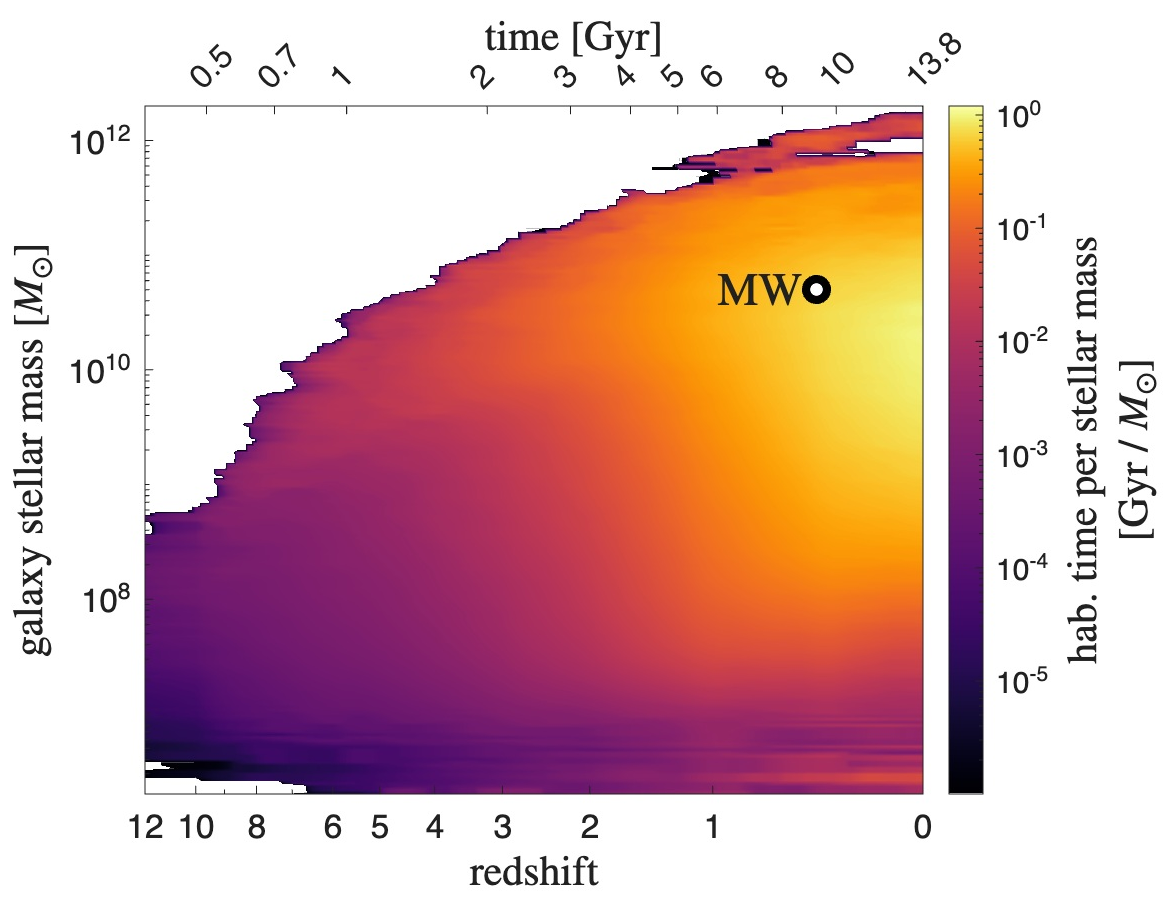}
	\end{minipage}
	\caption{The relationship between the habitable time per unit baryonic mass ($T_\ro{life,b}$, left) or per unit stellar mass ($T_\ro{life,s}$, right), within galaxies of given total halo mass (top) or stellar mass (bottom). The empty regions at the top left of each panel are a result of the finite simulation volume. Our 100~cMpc box size is too small to produce any galaxies in this region, and near its border we are subject to small number statistics. 				
		\emph{Top left}: Habitable time increases with cosmic time, as expected, and peaks for galaxies today in haloes with total mass in the range $10^{11} - 10^{12}~\Msol$. At this peak, the value $T_\ro{life,b} \sim 0.1$~Gyr/\Msol indicates that for every solar mass of baryons in a galaxy with total halo mass $\sim 10^{12}~\Msol$ today, 0.1~Gyr of time has passed on a habitable planet. At larger and smaller masses, galaxies are less efficient in turning baryons into habitable time. For most of cosmic history, galaxies with total mass about $10^{11} - 10^{13} \Msol$ are the most efficient at forming and preserving life-friendly environments. \emph{Top right}: this panel shows the habitable time per unit stellar mass, as a function of the total mass of the halo. Again, we see that galaxies with total mass about $10^{11} - 10^{13} \Msol$ are the most efficient at forming and preserving life-friendly environments. The \emph{bottom} panels show that this peak corresponds to galaxies with stellar masses of $10^{10} - 10^{11} \Msol$. }
	\label{fig:Nobs_vs_Mt}
\end{figure*}

\paragraph*{Gamma Ray Bursts:} GRBs are rarer than supernovae, but can affect life at much larger distances. In particular, the energy released is likely to be beamed into two narrow jets, each with half-radius ($\theta_b$) of a few degrees, corresponding to a fraction of the solid angle $f_b = 2~\Omega / 4\pi = \theta_b^2/2$ \citep{2001ApJ...562L..55F}. Most GRBs are believed to originate from core collapse in very massive (> 40 \Msol), rapidly rotating, low-metallicity stars \citep{2002ARA&A..40..137M}. Similarly to CCSN, we consider GRBs to be simultaneous with star formation.

We begin by assuming that the extinction radius of the GRB may be comparable to the SPH particle smoothing lengths. The rate of GRB extinction events as a function of time for a given star is,
\begin{equation} \label{eq:rGeInt}
r_\ro{Ge}(t) = \int_{<d_\ro{Ge}} f_b ~ \dot{n}_*(t,\bm{x}) ~ f_\ro{GRB}(Z) ~ \ro{d}^3\bm{x} ~,
\end{equation}
where the integral is over the region inside radius $d_\ro{Ge}$ around the star at time $t$, $\dot{n}_*(t,\bm{x})$ is the star formation rate density as a function of time and position, and $f_\ro{GRB}(Z)$ is the fraction of stars by number that will produce a GRB, which depends on metallicity. We perform this integral in our simulations by summing over the star formation rate for the gas particles around a given star particle,
\begin{equation} \label{eq:rGe}
r_\ro{Ge}(t) = f_b \sum_k f_{ik}^\ro{ol} ~ \dot{N}^*_{k} ~ f_\ro{GRB}(Z_k) ~,
\end{equation}
where $\dot{N}^*_{k}$ is the star formation rate for particle $k$, and $f_{ik}^\ro{ol}$ measures the fraction of the volume of the kernel of the $k$-th SPH particle that overlaps with a sphere of radius $R_\ro{Ge}$ around the $i$-th star particle.\footnote{Technically, we should be integrating the kernel over the overlapping region, not just comparing volumes. However, there is no convenient closed-form solution for this integral, and this level of precision is not required, given our other approximations. There is a formula for the volume: for two overlapping spheres of radius $R$ and $r$, with their centres separated by $d$, the volume of the overlap is $V = \pi (R + r - d)^2~(d^2 + 2d (r+R) - 3 (r^2+R^2) + 6rR)/(12d)$.} As above, for the razing probability, we simply substitute the razing radius into Equation \eqref{eq:rGe}, specifically the quantity $f_{ik}^\ro{ol}$.

In the case where the extinction or razing radius is smaller than the typical SPH smoothing length of the neighbouring particles, Equation \eqref{eq:rGeInt} becomes analogous to the case of CCSN,
\begin{equation}\label{eq:rGesmall}
r_{Ge} (t) = V_\ro{Ge} ~ f_b ~ \sum\limits_k f_\ro{GRB}(Z_k) ~ \dot{N}^*_k ~ W(\bm{x}_i - \bm{x}_k;h_k) ~,
\end{equation}
and similarly for razing ($V_\ro{Gr}$).

We model the dependence of the GRB fraction on metallicity using the ``GRB efficiency'' functional form of \citet{2013ApJ...773L..22T}, defined piecewise:
\begin{align}
f_\ro{GRB}(Z) &= f_0 ~ \frac{a \log_{10} Z/Z_\odot + b + p}{1 + p} \quad \ro{where} \\
(a,b) &=
 \begin{cases}
 (0,1), &{\color{white} 10^{-3} \leqslant} Z/Z_\odot \leqslant 10^{-3} \\
 (-3/8,-1/8), &10^{-3} \leqslant Z/Z_\odot \leqslant 10^{-1}\\
 (-1/4,0), &10^{-1} \leqslant Z/Z_\odot \leqslant 1\\
 (0,0), &{\color{white} 10^{-3} \leqslant} Z/Z_\odot \geqslant 1 ~.
 \end{cases}
\end{align}
This function is flat at small metallicities, and decreases as $Z$ increases. Following \citet{2015ApJ...802..103T}, we set $p = 0.04$. We infer the normalisation $f_0$ by ensuring that the total observed GRB rate in the simulation box of our Universe at the present day matches the observed global GRB rate, $10^9$ GRB/Gpc$^3$/Gyr \citep{2016PhRvL.116h1301P}. The observed rate includes the angular dependence, and so the combination $f_0~f_b$ is constrained by observation --- the value of $f_b$ only matters for the GRB extinction/razing radius (below). Given that the present-day star formation density is $\approx 10^{16}$ stars/Gpc$^3$/Gyr, the GRB fraction is $f_0 \approx 10^{-3}$.

Finally, we use an estimate of the beam angle from \citet{2017NatSR...7.5419S} as $\theta_b \approx 2^\circ$, which implies that $f_b = 1/1600$. This is toward the lower end of observational estimates, giving an upper limit on the implied extinction and razing radii \citep[see also][]{2001ApJ...562L..55F,2009ARA&A..47..567G,2015ApJ...815..102F}. From \citet{2001ApJ...562L..55F}, we assume a fiducial explosion energy of $10^{51}$ ergs; then Equation \eqref{eq:rEphi} tells us that a GRB will cause extinction at $d_{GRB,e} = 12$ pc and razing at $d_{GRB,r} = 0.4$ pc. These values are $1/\sqrt{f_b} \sm 40$ times larger than the corresponding SN radii, but still significantly smaller than the $20-50$kpc range used by \citet{2016PhRvL.116h1301P}. This is because they apply a fluence limit of $10^{7}-10^{8}$ which is relevant to Ordovician extinction-level events.

For the fiducial values in Table \ref{tab:params}, star-forming properties of the starburst galaxy M82 (10 $\Msol$/yr in the central nucleus of radius 0.5 kpc), and solar metallicity, $r_{Ge} \approx 4.3\ten{-3}~\ro{Gyr}^{-1}$. For a Milky-Way like galaxy, $r_{Ge} \sim 10^{-6}~\ro{Gyr}^{-1}$.

\paragraph*{Star-Star Interactions:} If another star wandered too close to the Solar System, the orbit of the Earth would be disrupted, potentially causing an extreme elliptical orbit or unbinding us entirely \citep{2022ApJ...935L...9B}. The stability of the orbits of planets requires that other stars do not perturb their orbits. Such an event is unlikely to cause a mere temporary extinction, so we will treat sufficiently large perturbations as razing events. 

The magnitude of the perturbation depends on the masses of the stars and planet, the semi-major axis of the planet, the velocity of the passing star, its impact parameter, and its angle of approach. We will use the Monte Carlo simulations of \citet{2015JCAP...09..030A}, who consider an Earth-mass planet in orbit around a Solar-mass star at 1 AU. They choose wandering stars from the stellar initial mass function with a Maxwell-Boltzmann type velocity distribution with width $v_b$. Averaging over the angle of approach, they give a fitting formula for the cross section for the interaction to cause orbital ellipticity $e > 0.5$ after the encounter,
\begin{equation}
	\sigma_\ro{dis}(v_b) \approx \frac{\sigma_0}{u(1+u)} \quad \ro{where} \quad u \equiv \frac{v_b}{1000 \kmsec} ~,
\end{equation} 
where $\sigma_0 = 1$ AU$^2$. The rate of disruptive encounters for a given star is,
\begin{equation}
	r_\ro{dis} = n_* ~ \sigma_\ro{dis}(v_b) ~ v_b ~,
\end{equation}
for local stellar density $n_*$ and velocity dispersion $v_b$. This rate is added to the total razing rate in Equation \eqref{eq:Pnr}. For $n_* = 1~\ro{pc}^{-3}$ and small velocities, $r_\ro{dis} \approx 2 \ten{-5}~\ro{Gyr}^{-1}$.

We use the local stellar density calculated in the previous section about SNIa. We calculate the local velocity dispersion (relative to particle $i$) by summing over the $n$ neighbours: $v^2_b = (1/n) \sum_j (v_j - v_i)^2$. We set the minimum velocity dispersion to be $v_b = 5 \kmsec$, which is typical for molecular clouds \citep{1991ApJ...378..565G} and dwarf spheroidal galaxies \citep{2007ApJ...667L..53W}. Because the cross section $v_b ~ \sigma_\ro{dis}(v_b)$ is approximately constant for small velocities, the rate of disruptive encounters is mostly insensitive to the velocity dispersion in small systems.

\paragraph*{Quasars:} 
We calculate the XUV luminosity from a black hole particle using the approach of \citet{2016MNRAS.462..190R}. Each BH particle records its current mass accretion rate, which is related to its bolometric luminosity as,
\begin{equation}
L_\ro{bol} = \epsilon_r ~ \dot{M}_\ro{accr} c^2 ~,
\end{equation}
where $\epsilon_r = 0.1$ is the radiative efficiency. The fraction of the total radiation that is emitted in the life-affecting XUV band is $\sm 20$\% \citep{2017MNRAS.464.1854B}, which we use to convert $L_\ro{bol}$ to $L_\ro{XUV}$. We track the cumulative fluence for star particles in the same halo as the black hole particle, and treat as uninhabitable any particles for which $\Phi_X > 10^{16.4}$ erg/cm$^2$. Because the impact of the quasar accumulates over $\sm$ millions of years, there is no obvious extinction level event that occurs for lower fluences. For example, the oceans will not boil if $5.6 \ten{33}$ erg of energy is absorbed in this way \citep{2017NatSR...7.5419S}, as unlike the aftermath of a nearby explosion, there is ample time for cooling to occur. Combining the equations gives, 
\begin{multline}
\Phi_X = 10^{16.5} \ro{erg/cm}^2  \left( \frac{\epsilon_r ~ \epsilon_{X}}{0.02} \right) ~\times~ \\
 \left( \frac{\dot{M}_\ro{accr}}{0.1 \Msol/\ro{yr}} \right) 
  \left( \frac{d}{300~\ro{pc}} \right)^{-2} 
  \left( \frac{t_{acc}}{100~\ro{Myr}} \right)  ~.
\end{multline}
Given that typical observed SMBH mass accretion rates are in the range $10^{-5}$ to 0.1~\Msol/yr \citep{2022ApJ...934...27S}, and quasar lifetimes are estimated to be $\sim 10~  \ro{Myr}$, the estimate above shows that the central regions of galaxies --- particularly more compact galaxies at high redshift --- can be substantially affected by QSO's.

We summarise the free parameters of our model in Table \ref{tab:params}, along with their fiducial values.

\section{Characterising the Galactic Habitable Zone (GHZ)}

\subsection{GHZ over cosmic history} \label{S:results}

In our model, the number of observers $N_\ro{obs}$ scales with the size of the box. We thus convert it into a density per unit mass or per unit comoving volume by dividing by the total mass or comoving volume of the simulation box.  $N_\ro{obs}(t)$ also depends on the constant but unknown probability ($f_l$) that life will form on a suitably habitable planet per unit time. As result, we will focus on the quantities,
\begin{equation}
T_\ro{life,b} = \frac{N_\ro{obs}(t)}{f_l ~ M_\ro{b,box}} \qquad, \qquad T_\ro{life,s} = \frac{N_\ro{obs}(t)}{f_l ~ M_\ro{s,box}}
\end{equation}
which gives the total amount of time that passes on habitable planets in the universe, per unit baryon mass (left) or per unit stellar mass (right). We will call this \emph{habitable time}. This is the amount of time that the universe creates for life, per unit mass. We will also be interested in the time derivative of these quantities, which we will call the \emph{habitable time rate}.

Figure \ref{fig:Nobs_vs_t} shows habitable time per unit baryonic mass (top panel) and per stellar mass (bottom), as a function of cosmic time (top axis) and redshift (bottom axis). Dashed lines show the predictions with no extinction events, while the solid line shows predictions will all extinctions. The habitable time can be interpreted as follows: a value of $1 \ten{-3} \ro{Gyr} / \Msol$ in the top (bottom) plot means that, on average, each solar mass of baryonic (stellar) matter at time $t$ has produced a million years of life-friendly time on suitable planets. 

Figure \ref{fig:dNobs_vs_t} shows the habitable time \emph{rate} per unit baryonic mass (top) and per stellar mass (bottom), calculated with (solid) and without (dashed) extinction events. This is the time derivative of Figure \ref{fig:Nobs_vs_t}.The habitable time rate can be interpreted as follows: a value of $1 \ten{-3} \Msol^{-1}$ in the top (bottom) plot means that, on average, for every billion years that passes and for each solar mass of baryonic (stellar) matter, a million years of life-friendly time passes on suitable planets.

The curves in Figures \ref{fig:Nobs_vs_t} and \ref{fig:dNobs_vs_t} rise rapidly in the early universe, driven by bursting star-formation and the rapid build-up of metals to make planets. In the late universe, once the universe has created the bulk of its stars --- but before a significant number of those stars have ended their lives --- the habitable time rate in the absence of extinctions (dashed curves) is roughly flat: each solar mass of baryons is creating about 7 Myr of life-friendly-planet time per Gyr, and solar mass of stars is creating about 100 Myr life-friendly-planet time per Gyr. In the distant future, as stars begin to die, these curves will begin to decrease.

Extinctions flatten the habitable time (equivalently, causes a downturn in the habitable time rate) at late times. Planets are being made uninhabited/uninhabitable faster than they are being produced.

\begin{figure}
	\centering
	\includegraphics[width=0.48\textwidth]{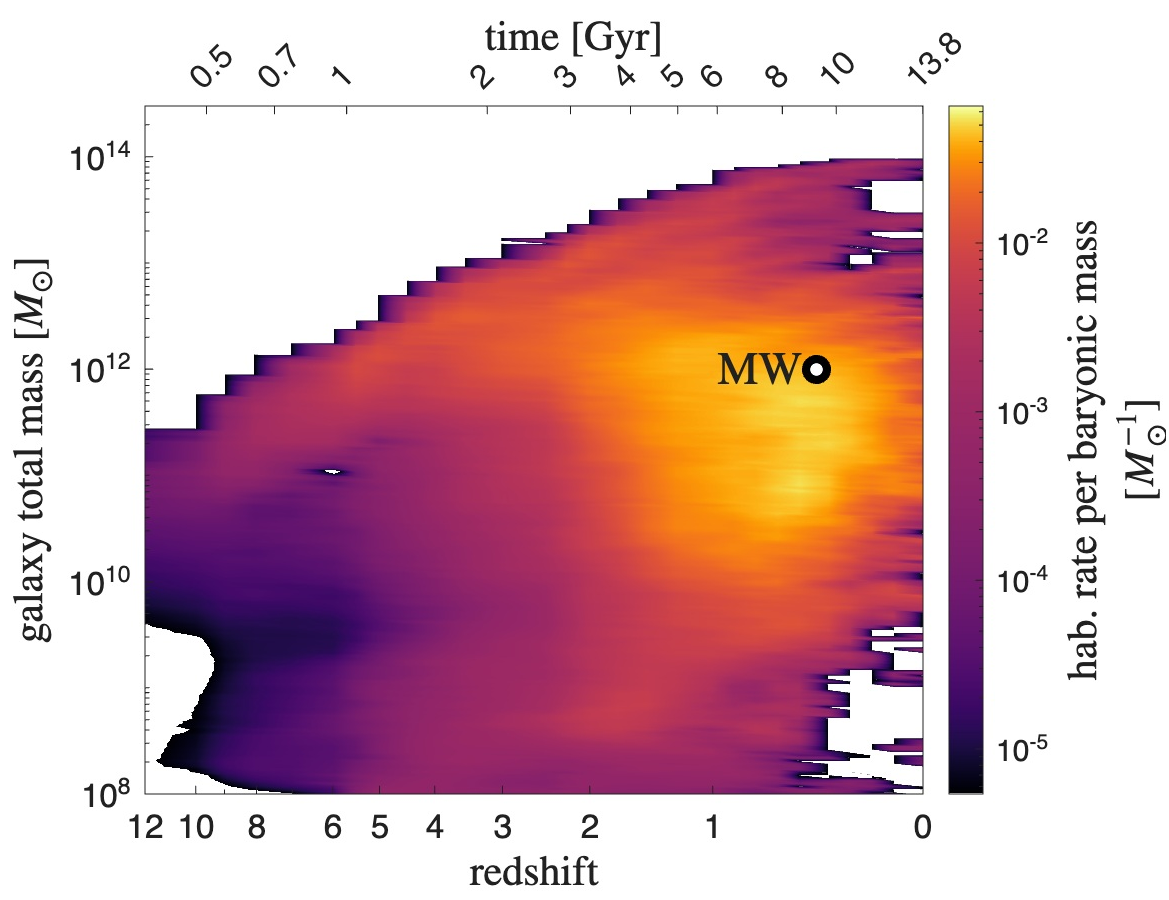}
	\caption{Habitable time \emph{rate} per unit baryonic mass, as a function of time and galaxy total mass. This is the time derivative of the top left plot in Figure \ref{fig:Nobs_vs_Mt}. While the numerical derivative makes the plot noisy, there is a decline in the rate at late times (the last few billion years).}
	\label{fig:Nobs_vs_dMt}
\end{figure}

\begin{figure*} \centering
	\begin{minipage}{0.48\textwidth}
		\includegraphics[width=\textwidth]{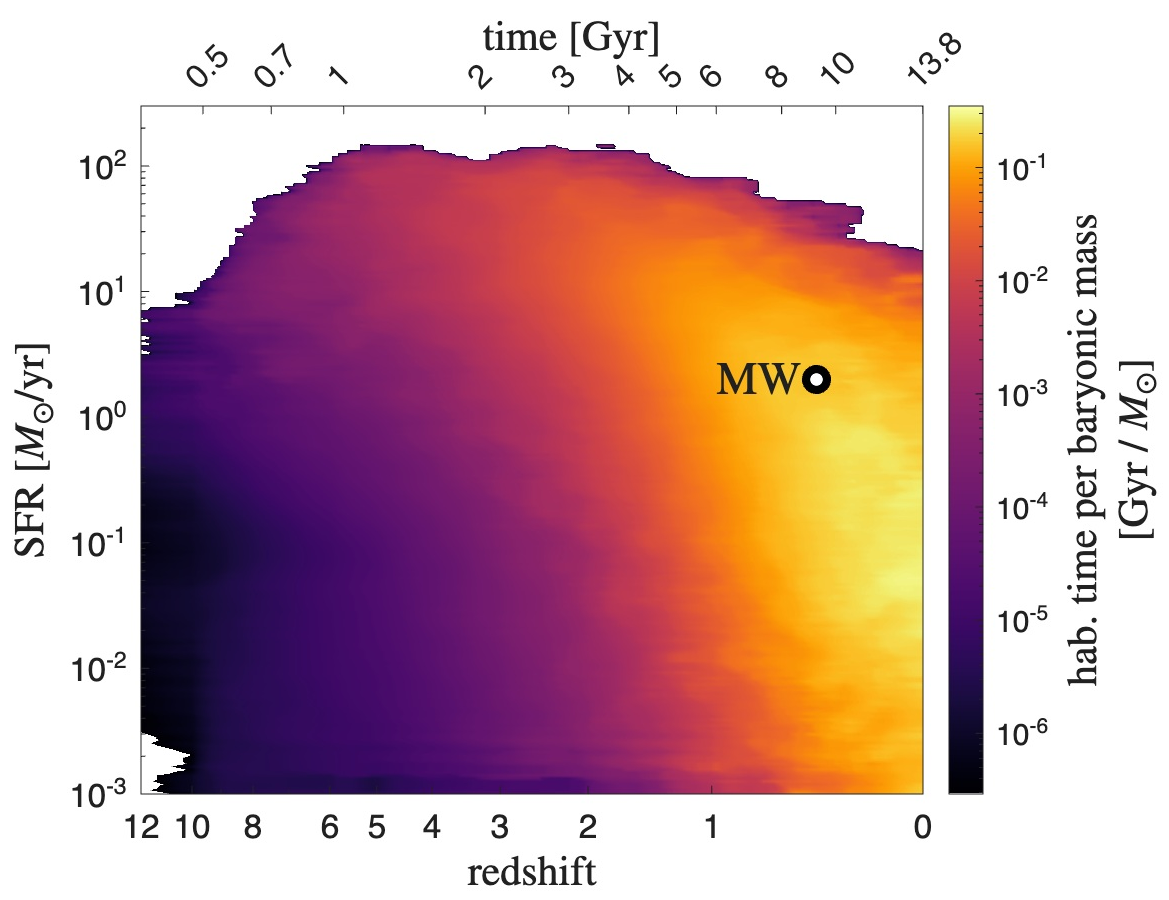}
	\end{minipage}
	\begin{minipage}{0.48\textwidth}
		\includegraphics[width=\textwidth]{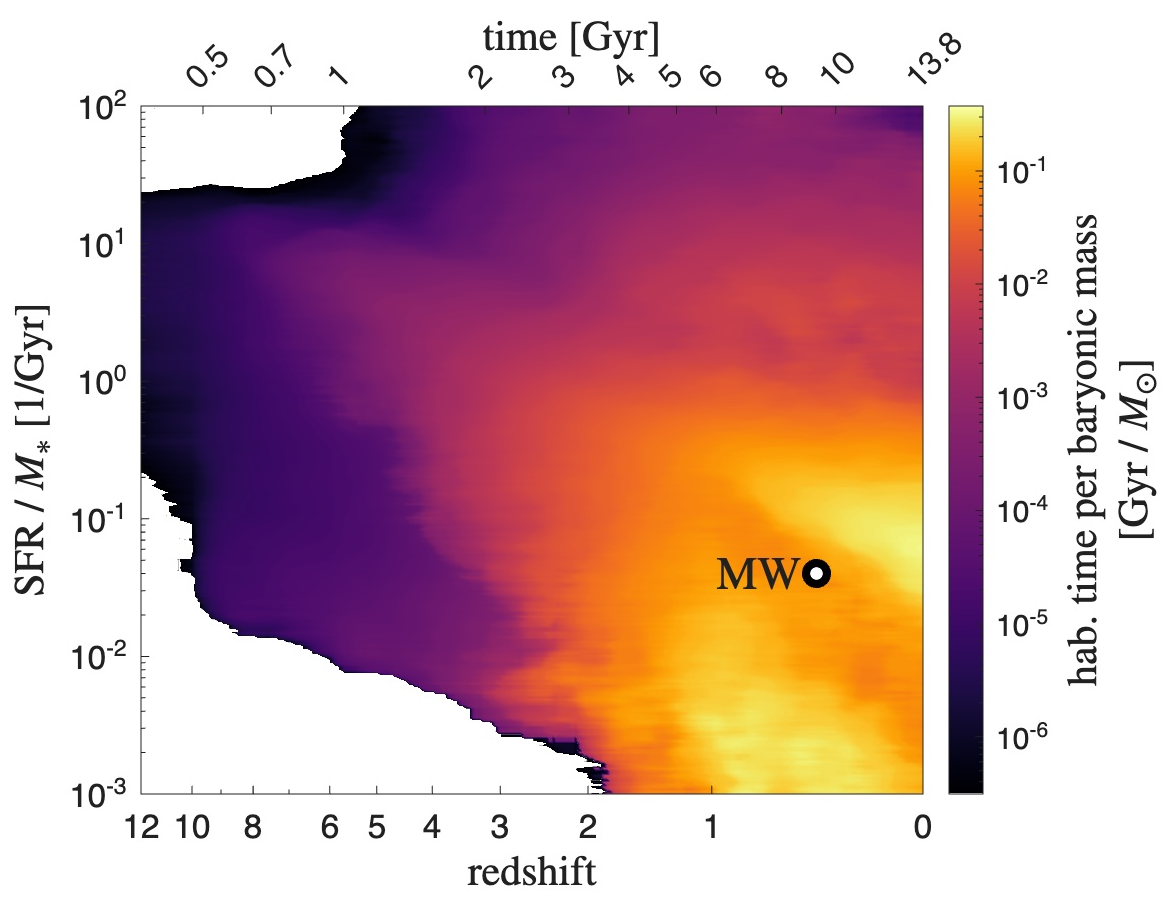}
	\end{minipage}
	\caption{The relationship between habitable time per unit baryonic mass and the star formation rate (SFR, left) and specific SFR (sSFR = SFR / $M_*$, right). The SFR \citep[$2 \, \Msol/$yr, ][]{2022ApJ...941..162E} and sSFR (0.04 Gyr$^{-1}$) of the Milky Way are labelled, at the redshift at which the Sun formed ($z_{\odot} = 0.42$).	The left panel shows that, before a redshift of $\sim 2$, habitable time was largest in the galaxies with the highest SFR. At later times, the largest habitable times are found in more quiescent galaxies. The right panel shows a similar trend, with the largest habitable times being found in galaxies with low sSFR in the late universe.} \label{fig:Nobs_vs_SFRt}
\end{figure*}

\begin{figure*} \centering
	\begin{minipage}{0.48\textwidth}
		\includegraphics[width=\textwidth]{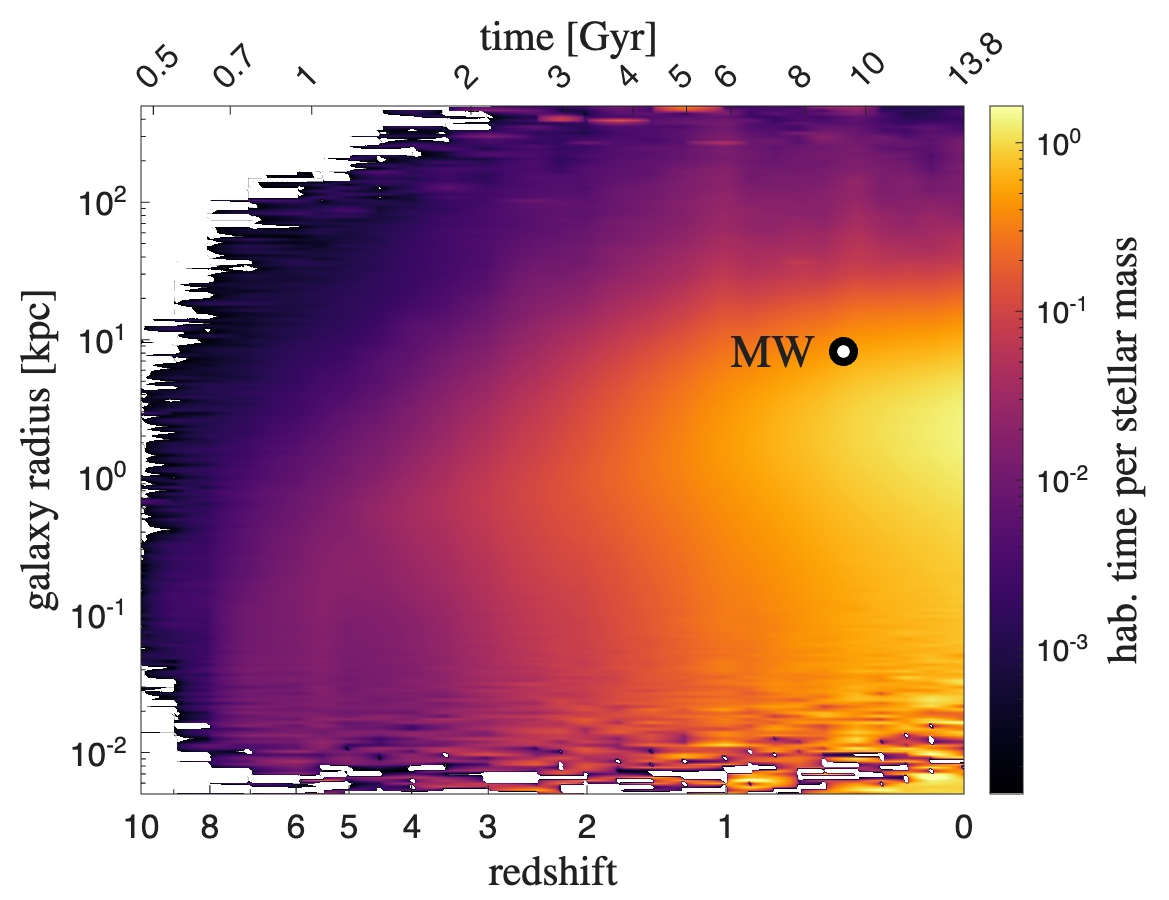}
	\end{minipage}
	\begin{minipage}{0.48\textwidth}
		\includegraphics[width=\textwidth]{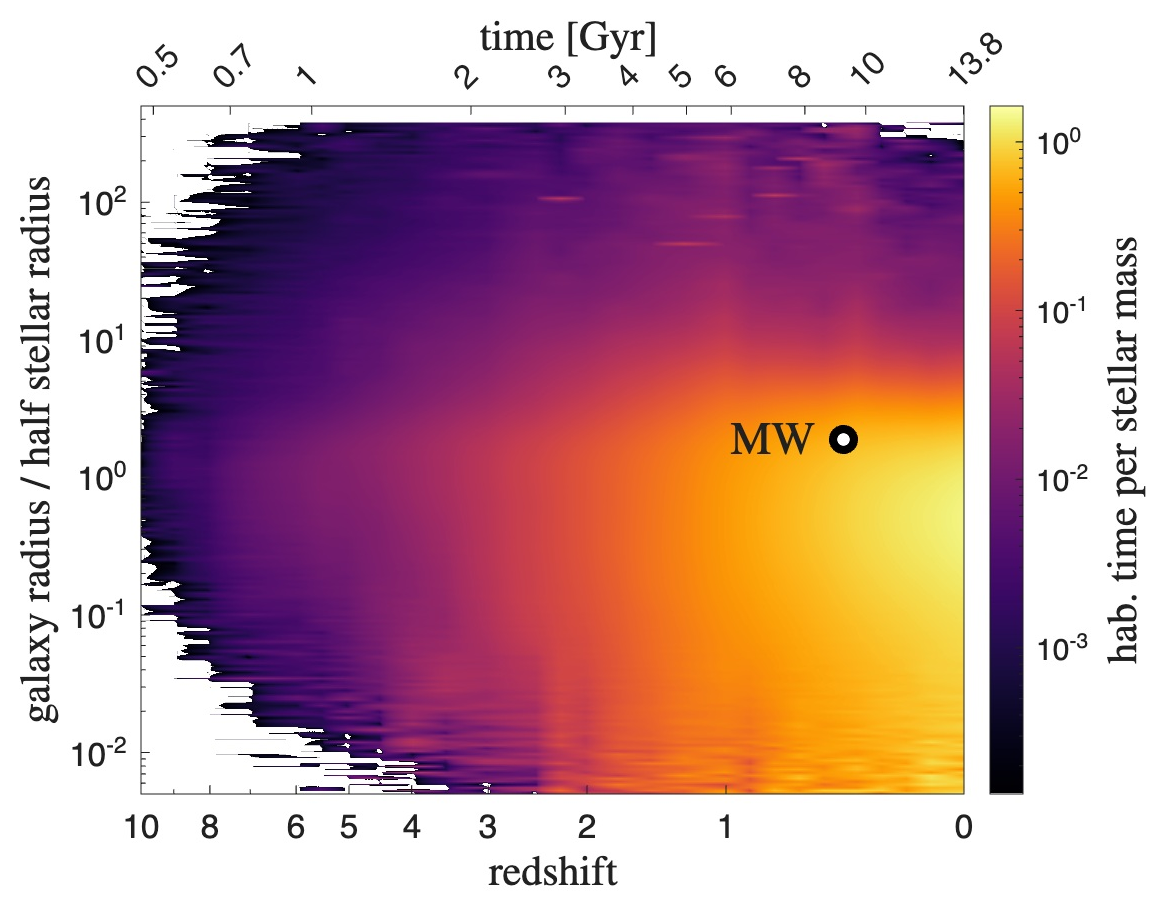}
	\end{minipage}
	\caption{The relationship between habitable time per unit baryonic mass (left) or per star (right), and distance of the star from the centre of its halo, for haloes with total mass in the range $10^{11} - 10^{13} \Msol$. The \emph{left} panel shows the relationship with the physical distance in kpc; the \emph{right} panel scales the physical distance for each star by $r_{50}$, the radius that contains 50\% of the stars in the halo. The label shows the Sun's galactocentric radius $8.2$~kpc and $r_{\odot} / r_{50} = 1.9$, at the redshift at which the Sun formed ($z_{\odot} = 0.42$).	
	The stars that are most efficient in creating habitable time are mostly found at radii from $\sim 1-10$ kpc, and $0.1 - 1 ~ r_{50}$ for most of the age of the history of the universe. The outer parts of galaxies are too metal poor to produce habitable planets, and the inner parts of galaxies are subject to the highest rates of extinction, as we will see below.} \label{fig:Nobs_vs_rt}
\end{figure*}

\subsection{Habitability, Galaxy Mass, and Star Formation Rate}
Which galaxies are most efficient at creating and preserving habitable planets? Figure \ref{fig:Nobs_vs_Mt} shows the relationship between the habitable time per unit baryonic mass ($T_\ro{life,b}$, left) or per unit stellar mass ($T_\ro{life,s}$, right), within galaxies of given total halo mass (top) or stellar mass (bottom).

Habitable time increases with cosmic time, indicating that planets are continuing to be formed around stars and are (for the most part) avoiding cosmic catastrophes. From the top panels, the habitable time peaks for galaxies in haloes with total mass in the range $10^{11} - 10^{12}~\Msol$. At this peak, the value $T_\ro{life,s} \sim 0.1$~Gyr/\Msol indicates that for every solar mass of baryons in a galaxy with total halo mass $\sim 10^{12}~\Msol$, 0.1$~$Gyr has passed on a life-friendly planet. For most of cosmic history, galaxies with total mass of about $10^{11} - 10^{13} \Msol$ are the most efficient at forming and preserving life-friendly environments. From the bottom panels, galaxies with stellar mass about $10^{10} - 10^{11} \Msol$ are the most efficient at forming and preserving life-friendly environments.

Figure \ref{fig:Nobs_vs_dMt} shows the habitable time \emph{rate} per baryonic mass, as a function of time and galaxy total mass. This is the time derivative of the top right plot in Figure \ref{fig:Nobs_vs_Mt}. Despite the noise, due to the numerical derivatives, the rate is seen to decrease at late times, indicating the effect of extinctions. 

In Figures \ref{fig:Nobs_vs_Mt} and \ref{fig:Nobs_vs_dMt}, the Milky Way is indicated with a circle, at the redshift at which the Sun formed ($z_{\odot} = 0.42$). The total (virial) mass is $M_{\ro{tot}} \approx 1 \times 10^{12} \Msol$ \citep{2012ApJ...761...98K,2019ApJ...873..118W} and stellar mass is $M_\ro{star} \approx 5 \times 10^{10} \Msol$ \citep{2015ApJ...806...96L,2016ARA&A..54..529B}. For comparison, \citet{2026arXiv260312317S} give the mass of the Milky Way at $z = 6$ as $M_{\ro{tot}} \approx 1 \times 10^{11} \Msol$, which is consistent with the model of \citet{2015MNRAS.450.1514C}.

The Milky Way sits in the high-mass end of the broad peak in habitable time per unit baryonic mass and per unit stellar mass. Habitable time accumulates with cosmic time, but habitable time rate (as we saw previously) seems to have peaked about 1Gyr ago.

Figure \ref{fig:Nobs_vs_SFRt} shows the relationship between habitable time per unit baryonic mass and the star formation rate (SFR, left) and specific SFR (sSFR = SFR / $M_*$, right). The left panel shows that, before a redshift of $\sim 2$, habitable time was largest in the galaxies with the highest SFR. At later times, the largest habitable times are found in more quiescent galaxies. The right panel shows a similar trend, with the largest habitable times being found in galaxies with low sSFR in the late universe.

\subsection{Habitability and Galactocentric Radius}

Figure \ref{fig:Nobs_vs_rt} shows the distribution of habitable time with galactocentric radius (in kpc, left) and  $r / r_{50}$ (right), which is the radius scaled to the radius that contains half of the mass of stars in the galaxy ($r_{50}$). The Sun's galactocentric radius is $8.2$~kpc \citep{2016ARA&A..54..529B}. The Milky Way's stellar half-mass radius is $4.25$~kpc \citep{2025ApJ...990L..37L}, so $r_{\odot} / r_{50} = 1.9$. These radii are indicated with a circle, at the redshift at which the Sun formed ($z_{\odot} = 0.42$).

There is a decline in habitable time at large radius because of a lack of metals to make planets. While there is some evidence of a decline at small radii, it is not particularly dramatic. The left plot shows a broad, stable late-time peak between 1 and 10~kpc (but recall that this averages over stars in galaxies of all masses.) In the right plot, the peak in habitable time per stellar mass shows a slight decline with scaled radius at late times, as planets find conditions closer to the centres of their galaxies more hospitable.

\section{Extinction Expectation}

\subsection{Relative Threat vs. Time}

\begin{figure} \centering
	\includegraphics[width=0.48\textwidth]{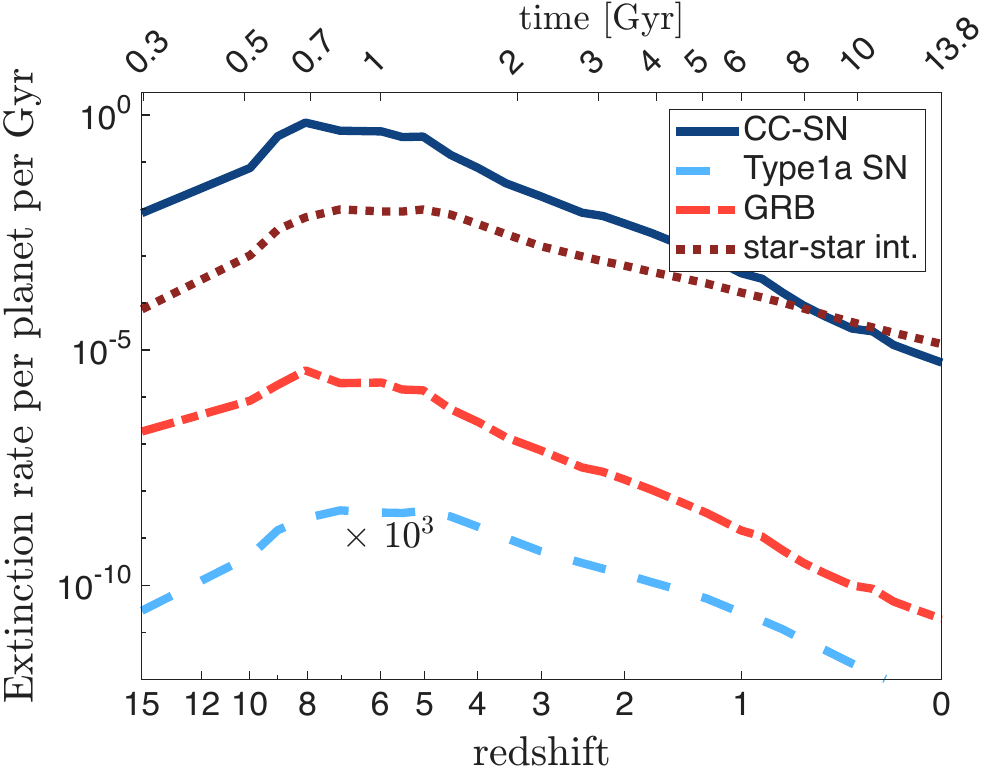}
	\caption{Extinction rate expectation value from each kind of cosmic extinction. Note that the Type 1a SN line has been multiplied by $10^3$. CCSN are the most likely source of extinction-level events in the early universe, while star-star interaction takes over in the last 5-6 Gyr.} \label{fig:absT_vs_tMtot}
\end{figure}

Which cosmic catastrophes are most significant for life in our universe? We calculate the expectation value of the extinction rate, averaged over all habitable planets at time $t$,
\begin{equation}
	\langle r_e \rangle = \frac{\sum_i r_{e,i} ~~ \frac{m^*_i}{\bar{m}_*} ~ f_\ro{ms} ~ P_\ro{HP}(Z_i)}{\sum_i \frac{m^*_i}{\bar{m}_*} ~ f_\ro{ms} ~ P_\ro{HP}(Z_i)} ~,
\end{equation}
where the sums are over all star particles; all symbols are the same as in Equation \eqref{eq:Nobs}. Note that razing events turn out to be rare, so we ignore them in this section. The results are shown in Figure \ref{fig:absT_vs_tMtot}. Note that the Type 1a SN line has been multiplied by 1,000. 

In the early universe, CCSN are the most likely source of extinction-level events, peaking at close to one event per Gyr between redshifts of $5 < z < 8$. GRBs, which also depend on the instantaneous star formation rate, follows a similar trend, but at a much lower late. For the last 5-6 Gyr star-star interactions have been more likely to cause an extinction, though at a relatively low rate $\lesssim 10^{-4}$~Gyr.

\subsection{Extinction, Galaxy Mass and Galactocentric Radius}

\begin{figure*} \centering
	\begin{minipage}{0.48\textwidth}
		\includegraphics[width=\textwidth]{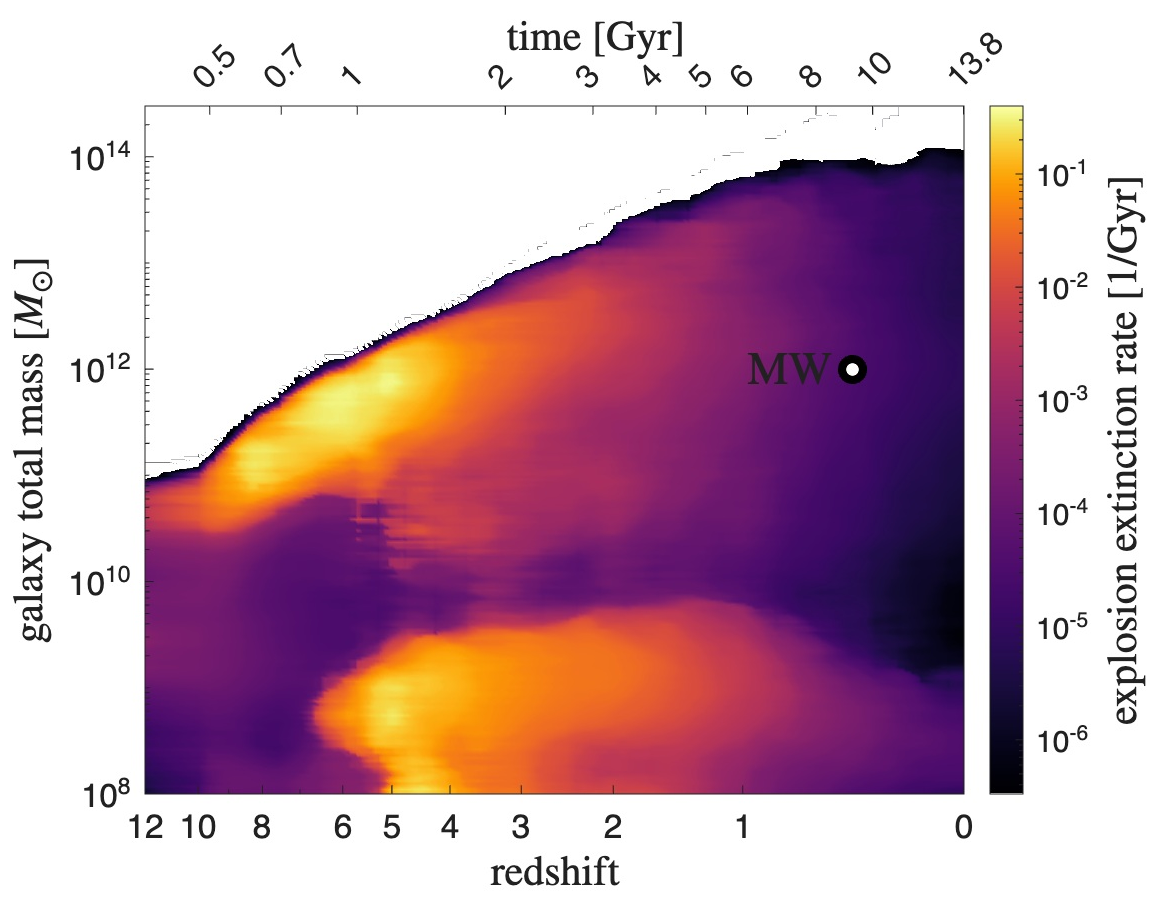}
	\end{minipage}
	\begin{minipage}{0.48\textwidth}
		\includegraphics[width=\textwidth]{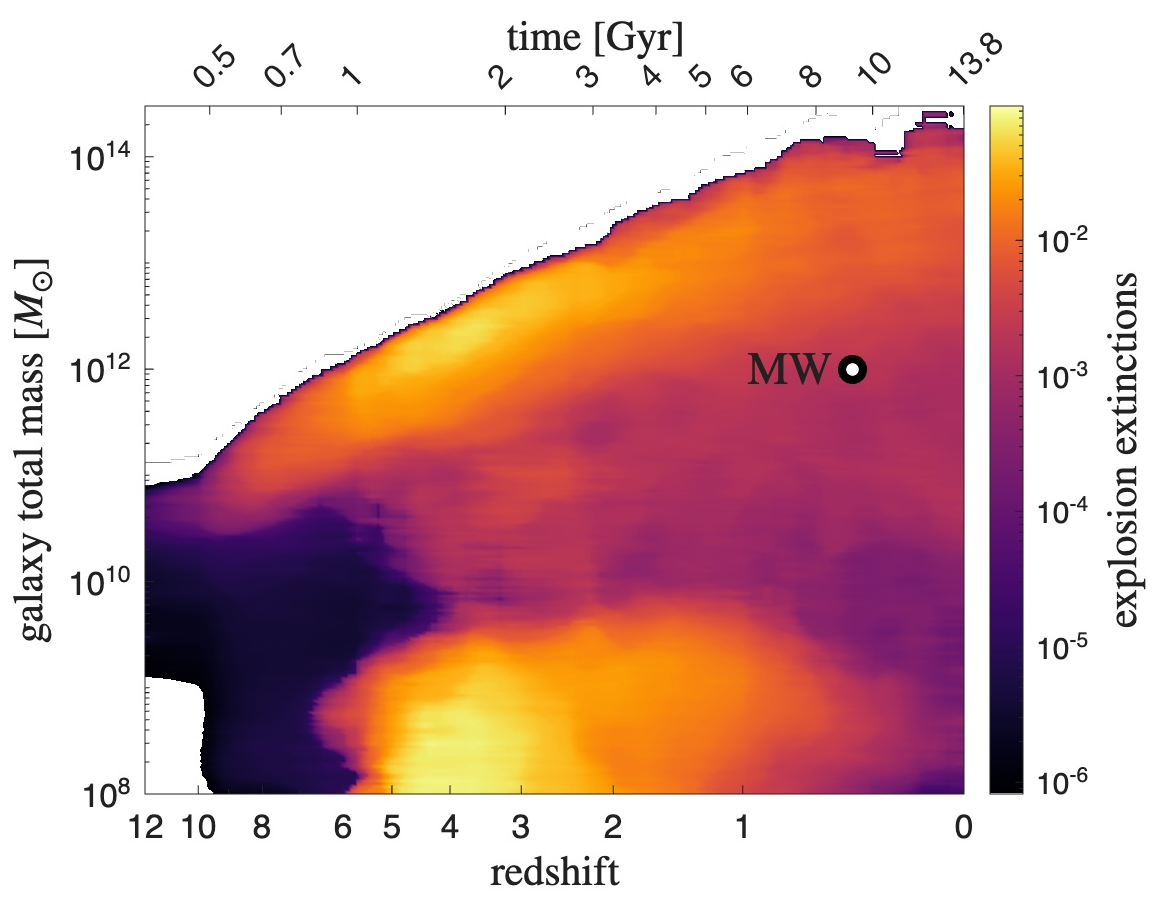}
	\end{minipage}
	\begin{minipage}{0.48\textwidth}
		\includegraphics[width=\textwidth]{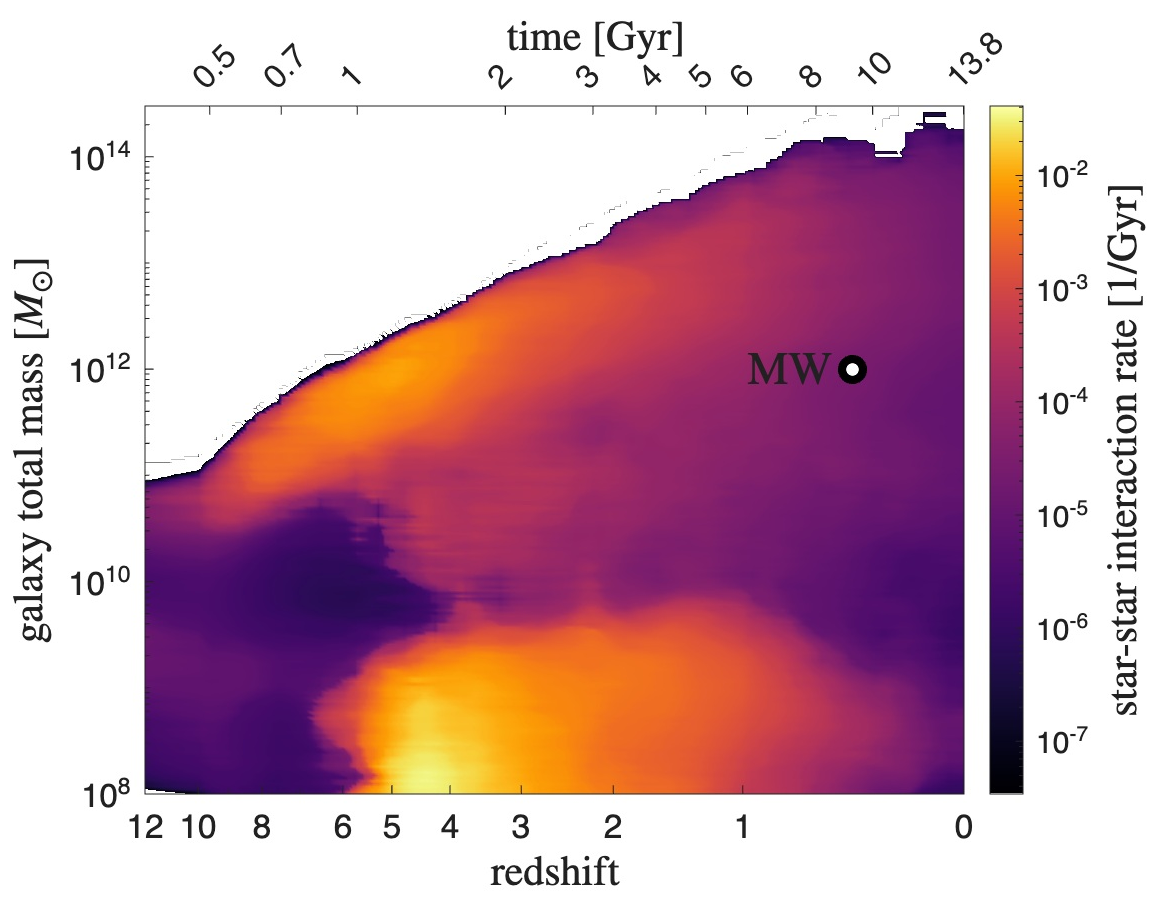}
	\end{minipage}
	\begin{minipage}{0.48\textwidth}
		\includegraphics[width=\textwidth]{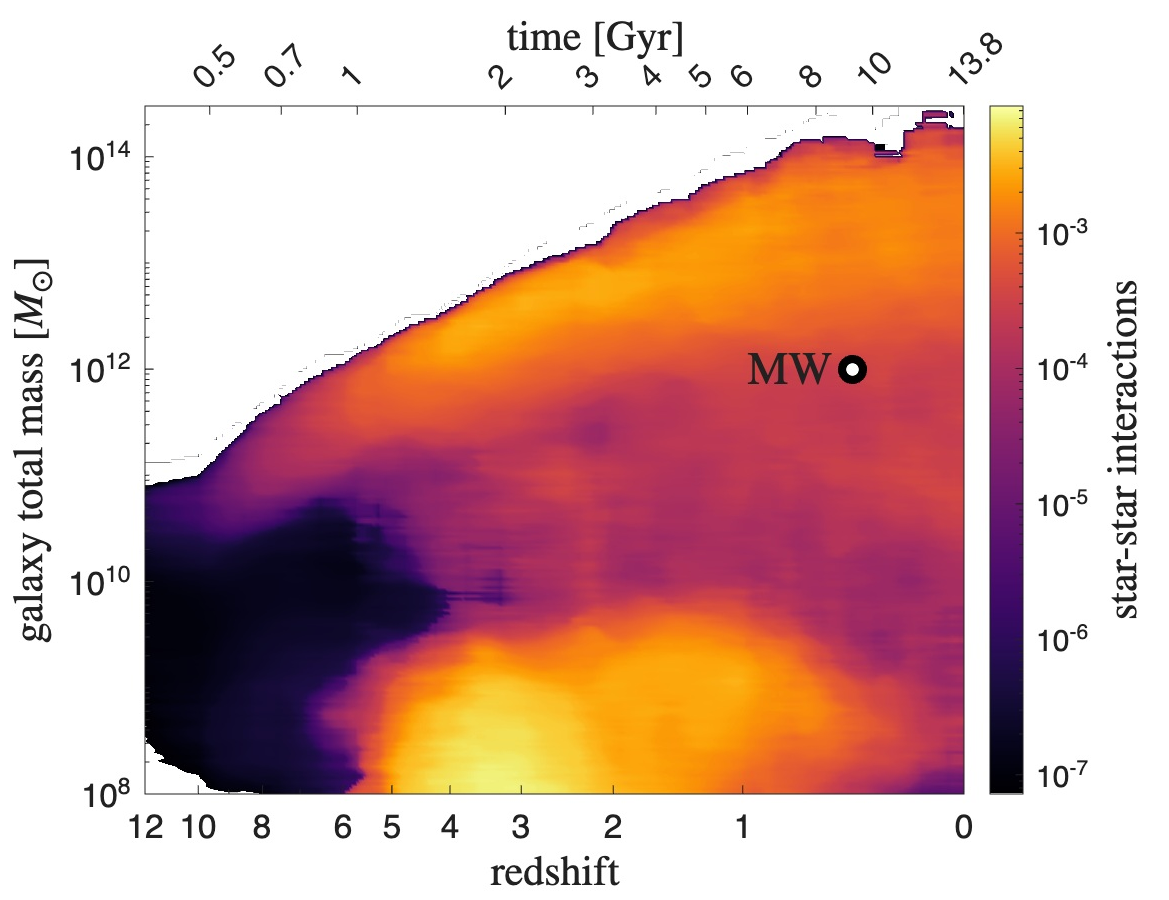}
	\end{minipage}
	\caption{The occurrance of extinction events in galaxies, as a function of their total mass. The top panels are for the explosive events (CCSN, Type 1a SN, and GRB), and the bottom panels are for star-star interactions. The left panels show the rate per Gyr, and the right plots show the expected number of events per star, integrated over the lifetime of each star. The plots show that newly-formed galaxies at redshifts $4 \lesssim z \lesssim 8$ (that is, about the first 1-2 billion years) are particularly dangerous. The first few billion years in a $\sim 10^{12} \Msol$ galaxy is the most dangerous place to be. These galaxies are dense and rapidly forming stars. Later, smaller galaxies of $\sim 10^{8} \Msol$ are also dangerous. In today's Milky Way, and without much change over the lifetime of the Sun (i.e. the last 4.5~Gyr), only about 0.1\% of stars would be expected to have experienced an extinction event by explosion or star-star interaction.} \label{fig:rallrss_vs_tMtot}
\end{figure*}

Figure \ref{fig:rallrss_vs_tMtot} shows the expectation value of (integrated) extinction rates as a function of the total galaxy mass. The top left panel is the expected rate for all explosive events (CCSN, Type 1a SN, and GRB). The top right panel shows the expectation value of the integrated explosion rate $R_{e,i} = \int r_{e,i} \dd t$. Note that while the value of $R_{e,i}$ for a given star cannot decrease with time, the expectation value can decrease as new stars are born. Figure \ref{fig:rallrss_vs_tMtot} (bottom row) is the expectation value of the rate and integrated rate for star-star interactions. 

The plots show that newly-formed galaxies at redshifts $4 \lesssim z \lesssim 8$ (that is, about the first 1-2 billion years) are particularly dangerous. For a $\sim 10^{12} \Msol$, the earlier the galaxy formed, the more dangerous it is, as galaxies that form earlier are more dense and form stars more rapidly. Later, smaller, compact galaxies of $\sim 10^{8} \Msol$ are also dangerous. In today's Milky Way, and without much change over the lifetime of the Sun (i.e. the last 4.5~Gyr), only about 0.1\% of stars would be expected to have experienced an extinction event by explosion or star-star interaction.

\begin{figure} \centering
	\includegraphics[width=0.45\textwidth]{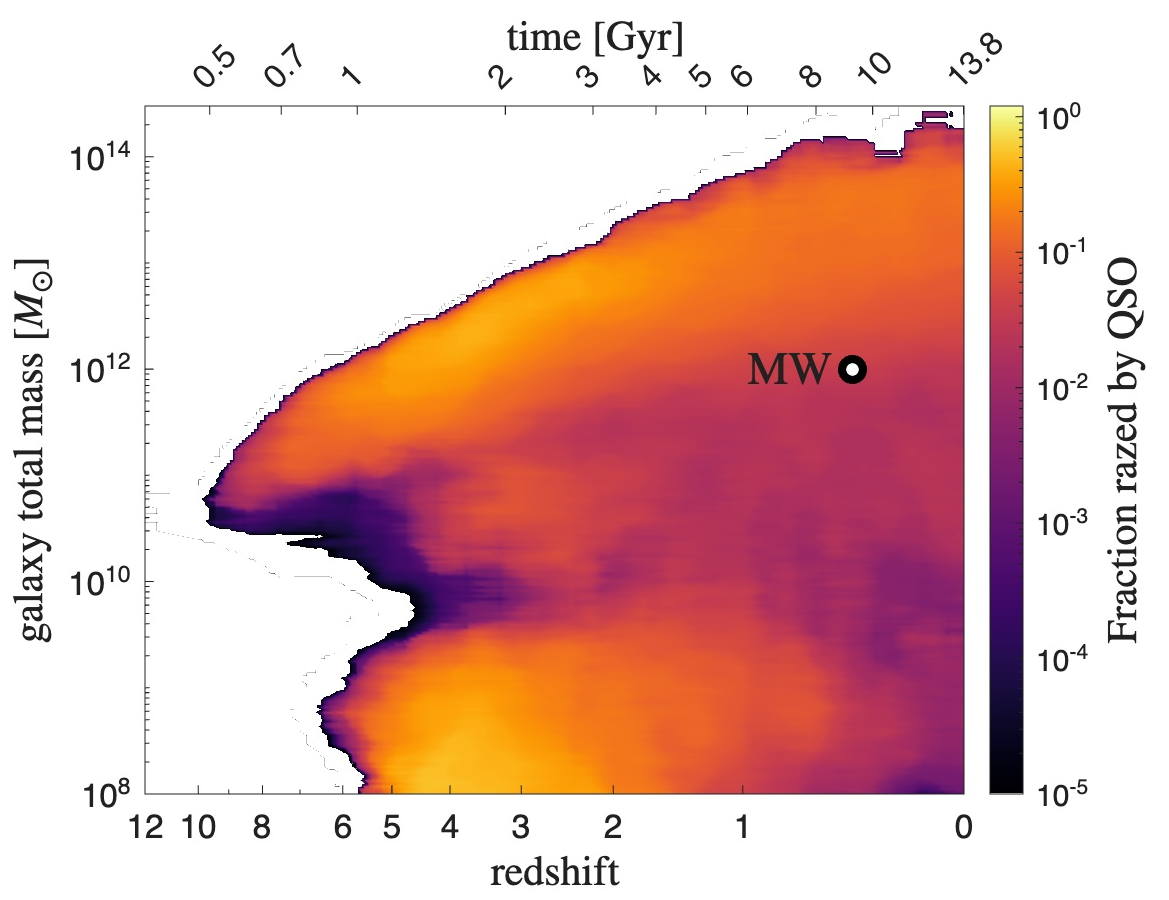}
	\caption{The fraction of habitable planets that have been razed by QSOs. As with other types of extinction, young newly-formed galaxies are the most dangerous. However, as discussed in the text, these fractions are surprisingly large.} \label{fig:BHdead_vs_tMtot}
\end{figure}

Figure \ref{fig:BHdead_vs_tMtot} shows the fraction of habitable planets that have been razed by QSOs. As with other types of extinction, young newly-formed galaxies are the most dangerous. Regardless, these fractions are surprisingly high: 5-30\% of all planets in $10^{12}-10^{13}~\Msol$ galaxies have had their atmosphere's stripped by QSOs. This may reflect an overestimate of the black hole accretion rate, or an overestimate of the QSO duty cycle due to the relatively long times between snapshots.

\begin{figure} \centering
		\includegraphics[width=0.45\textwidth]{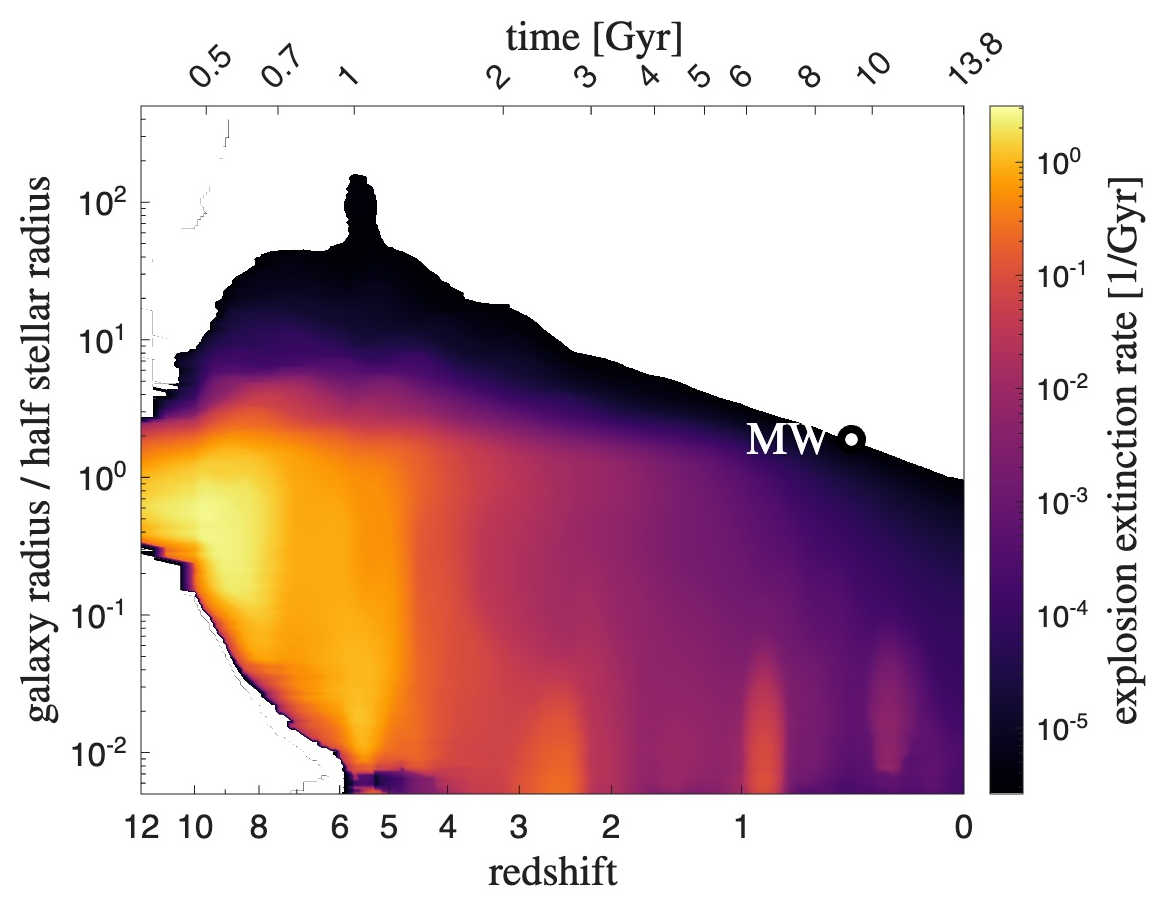}
		\includegraphics[width=0.45\textwidth]{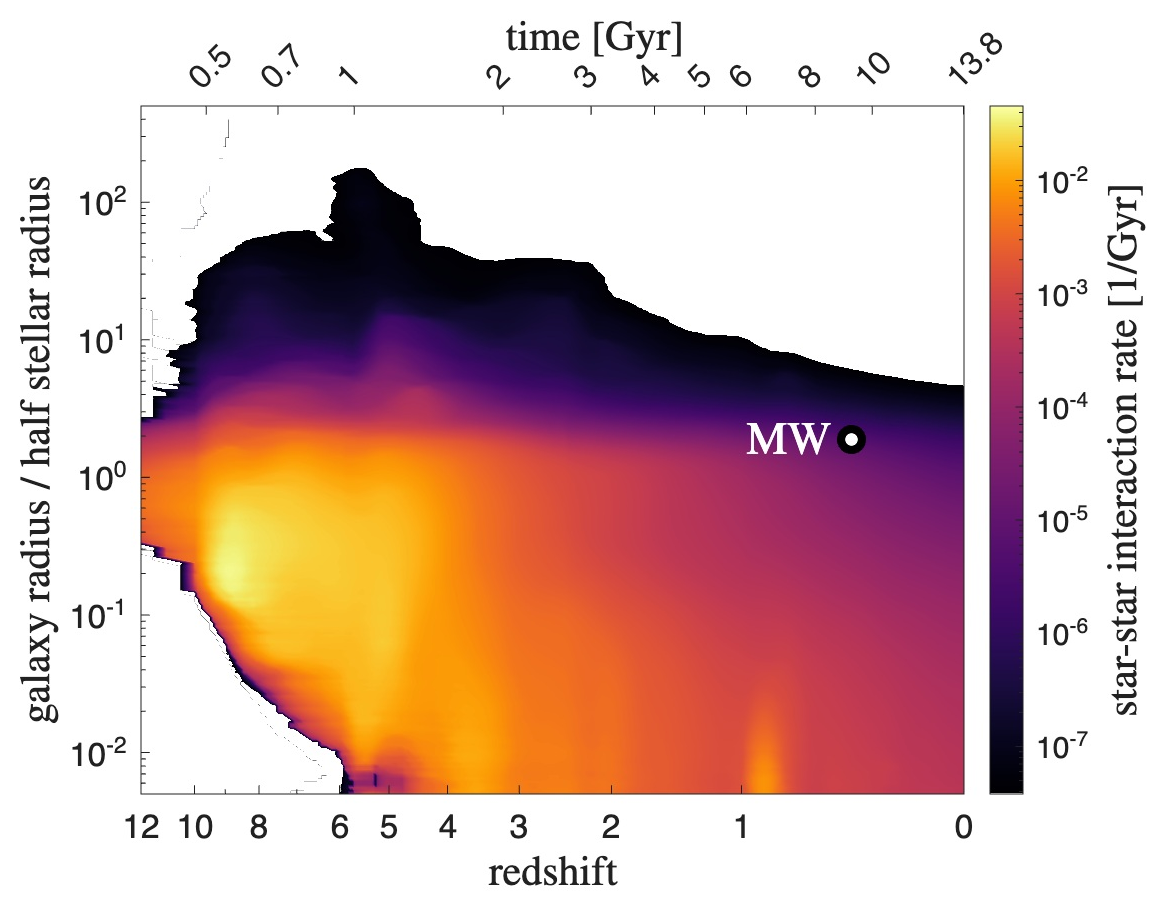}
	\caption{The rate of extinction events in galaxies, as a function of their scaled galactocentric radius and cosmic time. The top panel shows the rate of explosive events (CCSN, Type 1a SN, and GRB), and the bottom panel is for star-star interactions. We find that, as expected, the inner parts of galaxies are the most dangerous, with the peak of star formation in early dense galaxies leading to the highest extinction rates. Note that small radii calculations run up against limits from the resolution of the simulation.} \label{fig:rallrss_vs_tr50}
\end{figure}

Figure \ref{fig:rallrss_vs_tr50} shows the expected rate of extinction events, as a function of their galactocentric radius scaled to the half stellar radius, and cosmic time. As expected, the inner parts of the galaxy are more dangerous. Note that small radii calculations run up against limits from the resolution of the simulation.

\section{Conclusions}

We have presented a model for calculating the galactic habitable zone from a hydrodynamical galaxy formation simulation. The model takes into account star formation, stellar main-sequence lifetimes, habitable planet probabilities, and five sources of cosmic catastrophe: core-collapse supernovae, Type 1a Supernovae, gamma ray bursts, star-star interactions and quasar XUV radiation. Catastrophes are classified into extinctions (which make a planet uninhabited) and razing events (which render a planet uninhabitable). The number of observers depends on the probability of life forming on a planet, which is unknown. Thus, we focus on the habitable time created on planets per unit baryonic or stellar mass, and its time derivative (the habitable time rate).

With the fiducial parameters in Table \ref{tab:params}, we find that cosmic catastrophes cause the habitable time \emph{rate} to decline with time for the last few Gyr of cosmic history. The habitable time flattens as it approaches its current value of about 0.02 Gyr per solar mass of baryons, or 0.3Gyr per solar mass of stars. As a function of galaxy mass, habitable time peaks between about $10^{11}$ and $10^{12}$ \Msol total mass, whether measured per unit baryonic or stellar mass. Our galaxy, and our place within it, is typical of habitable planets. Habitable time does not show a particularly strong preference for galactocentric radius.

The most dangerous cosmic catastrophe in the early universe is core-collapse supernovae, while star-star interactions take over for the last 5 Gyr. Gamma ray bursts and Type 1a supernova are much less significant. The most dangerous galaxies are those that most efficiently form stars, during their initial burst of star formation. Smaller, compact galaxies of about $10^{8}$ and $10^{9}$ \Msol total mass are also dangerous. Only one star in a thousand in the present Milky Way can expect to have experienced an extinction due to a cosmic explosion.

While cosmological simulations allow for a more detailed and precise modelling of habitable zones, our conclusions are dependent on uncertain parameter values. For example, the rates of extinction and razing by cosmic explosions scale with their radii cubed (the volume), but are calculated via simple models. Our calculations here can be expanded and refined in a number of ways: more detailed models of local habitable zones and cosmic catastrophes, including the effects of cosmic rays, a higher cadence of simulation snapshots, greater resolution of galactic substructure (including stellar clusters and molecular clouds), and general improvements in simulation physics. We leave this for future contributions.

\section*{Acknowledgements} 
In memory of the late Prof. Richard Bower, who passed away in January 2023. His contributions to astronomy are many. His curiosity and humour are greatly missed.

Thanks to the referee for excellent comments.

We are grateful to all members of the Virgo Consortium and the \eagle collaboration who have contributed to the development of the codes and simulations used here, as well as to the people who helped with the analysis. This work was supported by the Science and Technology Facilities Council (grant number ST/P000541/1); European Research Council (grant number GA 267291 “Cosmiway”) and by the Netherlands Organisation for Scientific Research (NWO, VICS grant 639.043.409).

This work used the DiRAC Data Centric system at Durham University, operated by the Institute for Computational Cosmology on behalf of the STFC DiRAC HPC Facility (www.dirac.ac.uk). This equipment was funded by BIS National E-infrastructure capital grant ST/K00042X/1, STFC capital grants ST/H008519/1 and ST/K00087X/1, STFC DiRAC Operations grant  ST/K003267/1 and Durham University. DiRAC is part of the National E-Infrastructure. We acknowledge PRACE for awarding us access to the Curie machine based in France at TGCC, CEA, Bruyeres-le-Chatel.

\section*{Data Availability} 
The simiulation data of the EAGLE collaboration are available via \url{https://eagle.strw.leidenuniv.nl/}. Additional data underlying this article will be shared on reasonable request to the corresponding author.

\bibliographystyle{mnras}
\bibliography{2025FTGalSN}

\bsp 
\label{lastpage}
\end{document}